\documentclass[a4paper,11pt]{article}
\usepackage{amsmath}
\usepackage{amsfonts}
\usepackage{graphicx}
\usepackage{amssymb}
\usepackage{epsfig}
\usepackage{empheq}
\usepackage{hyperref}

\def\ba{\begin{eqnarray}}
\def\ea{\end{eqnarray}}
\def\be{\begin{equation}}
\def\ee{\end{equation}}

\def\nn{\nonumber}

\def\O{\mathcal{O}}

\def\red{\textcolor{black}}

\def\eps{\epsilon}
\def\L{\mathcal{L}}

\def\d{\partial}

\usepackage{tikz}
\usetikzlibrary{arrows,shapes}
\usetikzlibrary{trees}
\usetikzlibrary{matrix,arrows} 				
\usetikzlibrary{positioning}				
\usetikzlibrary{calc,through}				
\usetikzlibrary{decorations.pathreplacing}  
\usepackage{pgffor}							

\usetikzlibrary{decorations.pathmorphing}	
\usetikzlibrary{decorations.markings}

\numberwithin{equation}{section}

\begin{document}

\begin{titlepage}


Nikhef-2021-025

\vskip 1.5cm

\begin{center}
{\LARGE \bf{QFT without infinities and hierarchy problem}}
\vskip 1.5cm

\large{{\bf Sander Mooij$^{1,2}$ and Mikhail Shaposhnikov$^2$}}
\vskip 1cm

\emph{Nikhef, Theory Group \\Science Park 105, \\NL-1098 XG Amsterdam, The Netherlands}

\vskip 0.5cm

\emph{
Institute of Physics, Laboratory for Particle Physics and Cosmology,\\
	\'Ecole Polytechnique F\'ed\'erale de Lausanne, \\CH-1015 Lausanne, Switzerland
	}
	
\vskip 1cm

\tt{sander.mooij@epfl.ch,~mikhail.shaposhnikov@epfl.ch}

\end{center}

\vskip1cm

\begin{center}
\large{{\bf Abstract}}
\end{center}
\vskip 0.5cm


The standard way to do computations in Quantum Field Theory (QFT) often results in the requirement of dramatic cancellations between contributions induced by a  ``heavy'' sector into the physical observables of the ``light'' (or low energy) sector --  the phenomenon known as ``technical hierarchy problem''. This procedure uses divergent multi-loop Feynman diagrams, their regularisation to handle the UV divergences, and then re\-nor\-ma\-li\-sa\-tion to remove them. At the same time, the ultimate outcome of the renormalisation is the mapping of several  {\em finite} parameters defining the {\em renormalisable} field theory into different observables (e.g. all kinds of particle cross-sections). In this paper, we first demonstrate how to relate the parameters of the theory to observables without running into intermediate UV divergences. Then we go one step further: we show how in theories with different mass scales, all physics of the ``light" sector can be computed in a way that does not require dramatic cancellations induced by the physics of the ``heavy" sector. The existence of such a technique suggests that the ``hierarchy problem''  in renormalisable theories is not really physical, but rather an artefact of the conventional procedure to compute correlation functions. If the QFT is defined by the ``divergencies-free'' method all fine-tunings in theories with well-separated energy scales may be avoided.


\end{titlepage}

\newpage
\tableofcontents

\newpage

\section{Introduction}

\begin{figure}[h]
\centerline{\includegraphics[width=1\textwidth, angle = 0]{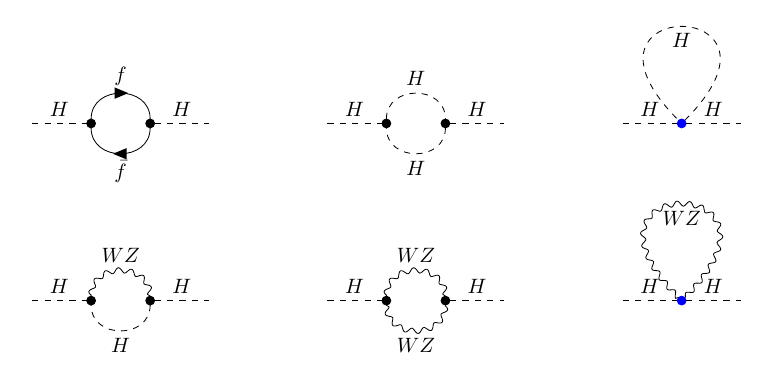}}
\caption{Higgs self-energy diagrams presented usually to explain the hierarchy problem.
} \label{selfenergy}
\end{figure}

One of the main directions which dominated the research in high energy physics is associated with the words  ``fine-tuning'' and ``naturalness''. In short, the puzzle can be formulated  as follows  (the discussion of these issues started from the works \cite{Gildener:1976ai,Weinberg:1975gm,Buras:1977yy,Susskind:1978ms} 
 for an incomplete set of reviews see, e.g. \cite{Susskind:1982mw,Haber:1984rc,Dvali:1995qt,Martin:1997ns,Chung:2003fi,Giudice:2008bi,Feng:2013pwa,Dine:2015xga,Nath:2020xiz,Hebecker:2021egx}). Let us take the Standard Model (SM) and consider radiative corrections to the Higgs mass. The contributions come from the top, Higgs and gauge loops, see Fig. \ref{selfenergy}. Several of these diagrams are quadratically divergent, leading to the most significant term $\delta m_H^2 \propto f_t^2\Lambda^2$, where $f_t$ is the top quark Yukawa coupling and $\Lambda$ is the ultraviolet cutoff of the theory, i.e. the place where the Standard Model is substituted by the more fundamental theory of Nature. Thus, if the physical value of the Higgs boson mass  $m_H$ is small compared to $\Lambda$, $m_H \ll \Lambda$, one has to fine-tune the tree mass $M_{\mbox{tree}}$ to cancel the radiative correction(s). The ``amount of fine-tuning'' is usually defined by \cite{Barbieri:1987fn,Contino:2017moj,deBlas:2019rxi,Strategy:2019vxc}
\be
\epsilon_H=\frac{M_{\mbox{tree}}^2-\delta m_H^2}{\Lambda^2}\sim\left( \frac{100~\mbox{GeV}}{4\pi \Lambda}\right)^2~.
\label{epsH}
\ee
If $\epsilon \sim 1$ the theory is considered to be ``natural'', and if $\epsilon \ll 1$  -- ``unnatural'' \cite{Barbieri:1987fn}, underlying that the cancellation between two large numbers looks very improbable. This is the so-called problem of quantum stability of the electroweak scale against quantum corrections\footnote{The stand-alone Standard Model is a theory with just one physically relevant mass parameter - the Fermi scale - and thus it does not have fine-tuning issues if the Landau poles in the Higgs self-coupling and hypercharge gauge couplings are disregarded.}. 

A similar logic can be applied to another class of badly (quartically) divergent diagrams in QFT, defining the vacuum energy $\epsilon_{\mbox {vac}}$. Here the radiative corrections are proportional to the fourth power of the cutoff scale,  $\delta \epsilon_{\mbox {vac} }\propto \Lambda^4$ leading to an even higher degree of fine-tuning in the cosmological constant: 
\be
\epsilon_{\mbox{cc}}=\frac{\epsilon_{\mbox {vac}}^{\mbox{tree}}-\delta \epsilon_{\mbox {vac}}}{\Lambda^4}\sim \left( \frac{10^{-3}~{\mbox{eV}}}{\Lambda}\right)^4~.
\label{epsvac}
\ee

An equivalent picture arises in the Wilsonian approach of going from high energy theory to effective theories defined at low energies. When defining the low energy description of Nature provided by the SM, one considers all sorts of gauge-invariant operators  ${\cal O}_n$ of mass dimension $n$, constructed from the SM fields.  As the Lagrangian has mass dimension four, simple power counting reveals that only two operators in the expansion of the action with respect to possible operators may come with positive powers of the cutoff, namely ${\cal O}_2 \propto \Lambda^2 h^\dagger h$, giving the mass of the Higgs boson ($h$ is the scalar field of the SM), and  ${\cal O}_0 =\epsilon_{\mbox {vac}}\propto \Lambda^4$, representing the vacuum energy. The so-called fine-tuning puzzle is why the high-energy contributions to these quantities are nearly cancelled by the low-energy radiative corrections, leading again to problems with naturalness. The operators of dimension four constitute the Lagrangian of the SM, whereas higher-dimensional operators, suppressed by several powers of the cutoff scale $\Lambda$, lead to different effects such as neutrino masses or proton decay \cite{Weinberg:1979sa,Wilczek:1979hc}. 

What is the value of the cutoff $\Lambda$? With the parameters fixed by the experiment (most notably the masses of the top quark and the Higgs boson) the SM by itself is valid up to exponentially high energies, $\Lambda \propto M_W \exp{(-c/\alpha)}$ (here $\alpha$ is a typical value of the coupling constants in the SM and $c$ is some numerical constant). The cutoff is associated with Landau poles in the Higgs boson self-coupling and the hyper-charge gauge coupling constants\footnote{Note that the experimental drawbacks of the SM: neutrino masses and oscillations, the presence of non-baryonic Dark Matter, and baryon asymmetry of the Universe provide no hint about the SM cutoff. In particular, all these problems can be solved by adding to the SM three Majorana fermions with masses below the Fermi scale \cite{Asaka:2005pn}. Inflation can be driven by the Higgs boson of the SM and does not require introducing any new particle \cite{Bezrukov:2007ep}.}. Gravity, yet another force of Nature, enters at a much lower scale than the Landau poles, $M_P\sim 10^{19}$ GeV. (For a discussion of the hierarchy problem and black holes, see \cite{dva3}.) If the strong, weak and electromagnetic interactions are part of some Grand Unified Theory (GUT), then $\Lambda \sim 10^{15-16}$ GeV \cite{Georgi:1974yf}. With the use of these values for the cutoffs, one concludes from (\ref{epsH}) that the SM is highly fine-tuned and therefore ``unnatural''.  The severity of the fine-tuning of the cosmological constant associated with the  $\epsilon_{\mbox{cc}}$  is striking. (However, in \cite{hossi} it was argued that in the absence of a physically motivated measure to decide whether a number is ``unnatural or not", there is no reason to worry about any value for $\eps_H$. References \cite{wells1,wells2} propose a ``moderate naturalness position", arguing that even in the absence of such a measure, the search for ``natural theories" can still be defended.)

On the experimental side, several bounds have been established on the scale of new physics beyond the SM (see, e.g. \cite{Ellis:2021kzk}), which can be summarized as constraints on coefficients in front of different higher-dimensional operators appearing in SMEFT (SM Effective Field Theory, for a review, see \cite{Brivio:2017vri}). Depending on the channel\footnote{There are few 3$\sigma$ signals that may indicate the presence of new physics coming from LHCb experiments \cite{LHCb:2021trn} and from the anomalous muon magnetic moment \cite{Muong-2:2021ojo}. It remains to be seen whether they will lead to discovery.}, $\Lambda>{\cal O}(10-100)$ TeV.  According to the criterion (\ref{epsH}), this means that the SM is an unnaturally fine-tuned theory with $\epsilon_H \lesssim 10^{-2}$. 

\begin{figure}[h]
\centerline{\includegraphics[width=0.6\textwidth, angle = 270]{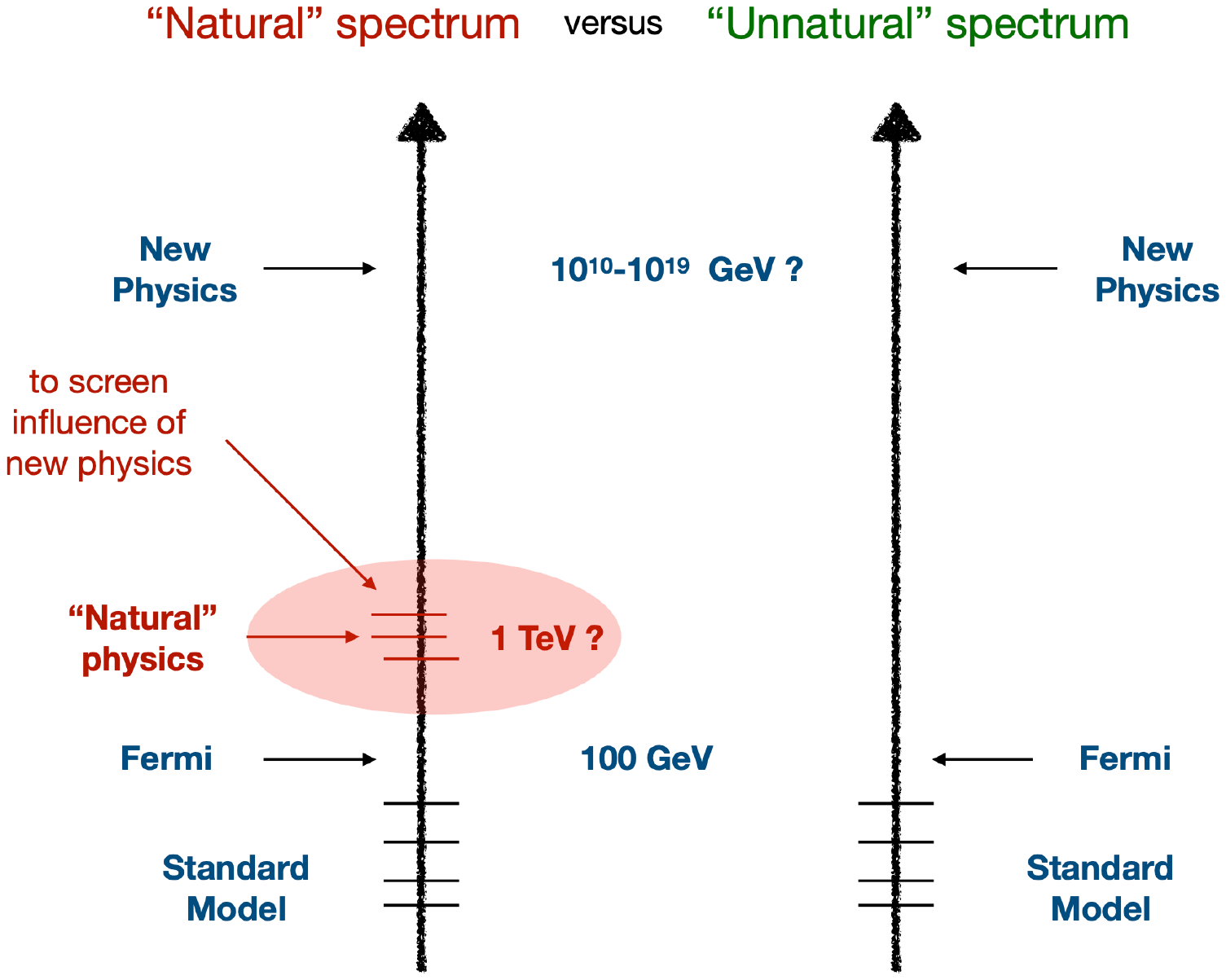}}
\caption{The common naturalness argument caught in a picture. At the left, ``natural" (TeV?) physics lowers the cut-off of the Standard Model. Only a mild tuning between loop corrections and counterterms is required to explain the apparent stability of the Higgs mass. At the right, the Standard Model holds up to much higher energy scales. A dramatic fine-tuning is needed to keep the Higgs field light: if there is no TeV physics, how can this tuning be maintained?} \label{plaatjemisha}
\end{figure}

These arguments about fine-tunings and naturalness seem to be so convincing, that they are motivated to search for ``natural'' extensions of the SM.  The paradigm is based on the idea that the {\bf low energy physics}  should be organised in such a way that the small value of the electroweak scale should be stable against modifications of the ultraviolet (UV) physics and thus does not depend on the Grand Unified or Planck scales. This naturalness criterion necessitates some kind of  ``Beyond Standard Model" (BSM) physics right above the Fermi scale $G_F^{-1/2}$ lowering the SM cutoff to $\Lambda \sim G_F^{-1/2}\sim {\cal O}(100)$ GeV and thus making the SM ``natural'' (for review see, e.g.  \cite{Giudice:2008bi,Giudice:2013yca}).  This new physics is supposed to ``screen'' the large UV scale from the low energy domain, see Fig.  \ref{plaatjemisha}. The examples include -- but are not limited to -- low energy supersymmetry, composite Higgs boson, or large extra dimensions (for reviews see, e.g. \cite{ParticleDataGroup:2020ssz}).  Yet other more recent suggestions are that the fine-tunings may result from cosmological evolution \cite{dva1,dva2,Graham:2015cka} or environmental selection \cite{Giudice:2021viw}.  In general, these proposals also lead to the prediction of new (scalar) particles around the Fermi scale \cite{Banerjee:2020kww}.

The spectra of elementary particles are drastically different in ``natural'' and ``unnatural''  theories.  Since the strategy of how to search experimentally for new physics depends on its scale, one should understand how robust are the naturalness arguments.  This paper aims to analyse the fine-tunings visibly appearing when quantum corrections to the Higgs mass or the cosmological constant are included.

This problem -- the alleged fine-tuning between loop corrections and counter-terms -- is sometimes referred to as ``the technical hierarchy problem", or as just "the hierarchy problem". In this paper, we will use the term ``technical hierarchy problem". We need to be very clear about this, since there is yet another, firmly related, alleged problem that also goes under the name ``hierarchy problem." This problem appears when one tries to compute the Higgs mass from a hypothetical UV complete theory having the Standard Model as an infrared limit. In a top-down computation of the Higgs mass from that theory, a precise tuning between the parameters of the UV theory is claimed to be needed to reproduce the small Higgs mass. We will \red{present our personal opinion} on this problem in Section \ref{calc}, and in the Conclusions. Apart from that, we will always be concerned with the first problem, and refer to it as ``the technical hierarchy problem".\footnote{A third formulation of the hierarchy problem addresses the puzzling fact that the physical Higgs mass or the cosmological constant is so small compared with the scale of new physics (such as gravity or hypothetical Grand Unification). This problem is not addressed at all in our work.}

In short, in this paper, we will examine whether the ``fine-tunings''  discussed above represent a real physical problem or are related to the commonly used formalism of QFT based on Feynman graphs and renormalisation.  Of course, one can  wonder:  ``Is there anything else?'' 

Presently, there is a revival of interest in non-perturbative formulations of QFT. Besides an obvious situation of the strong-coupling regime when the perturbation theory does not work, the search for alternatives is important from a conceptual point of view -- for instance, it is far from clear how to include gravity in the overall picture. Several examples of recent progress include the direct computation of the S-matrix amplitudes (summing up contributions of many diagrams) based on generic principles of QFT such as unitarity and Lorentz-invariance. This has proven to be a very efficient and conceptually clean method (for a review see, e.g. \cite{Cheung:2017pzi}). The non-perturbative conformal bootstrap \cite{Rattazzi:2008pe,El-Showk:2012cjh} made a breakthrough in the computation of anomalous dimensions of different operators and critical indices in the theory of phase transitions. The S-matrix bootstrap \cite{Paulos:2016fap} allowed to derive strict bounds on the values of different couplings. 

Naturally, the question arises: can we have a formulation of QFT in which there are no fine-tunings in theories containing very different physical mass scales?

In a search for a QFT formulation without fine-tunings, one should identify the very starting point leading to the apparent cancellation of very large numbers, namely the ``tree'' Higgs mass and its qua\-dra\-ti\-cal\-ly divergent radiative corrections. Clearly, the devil sits in the quadratic (or quartic, if we talk about the cosmological constant) divergences, inevitably appearing in Feynman diagrams with loops in theories with fundamental scalar fields. 

Let us look at this point more closely. During several decades of the 20th century, the research in quantum theory was concentrated on the struggle with UV infinities that appear in perturbation theory. This culminated in the creation of the renormalisation procedure in QFT. Namely, it happened to be possible to formulate the set of rules in which the UV divergent expressions are first regularised,  so that Feynman integrals acquire mathematical sense, and then the divergences are subtracted with the use of so-called counter-terms --  UV-singular operators which are added to the Lagrangian of the theory (a relevant textbook is, for instance, \cite{Collins:1984xc}).  In the renormalisable theories (like the SM) there is only a finite number of counter-term structures, which are needed to remove divergences. This leads to theories in which the results of all experiments can be predicted, i.e. expressed through a {\em finite} number of parameters\footnote{Or, in experimental terms: from the results of few experiments we can predict the outcome of infinitely many experiments.} (such as the mass and the charge of the electron when we consider quantum electrodynamics).  In non-renormalisable theories (like gravity) an infinite number of counter-term structures is required to remove the divergences, leading to an infinite number of parameters characterising the theory.

\begin{figure}[h]
\centerline{\includegraphics[width=1\textwidth, angle = 0]{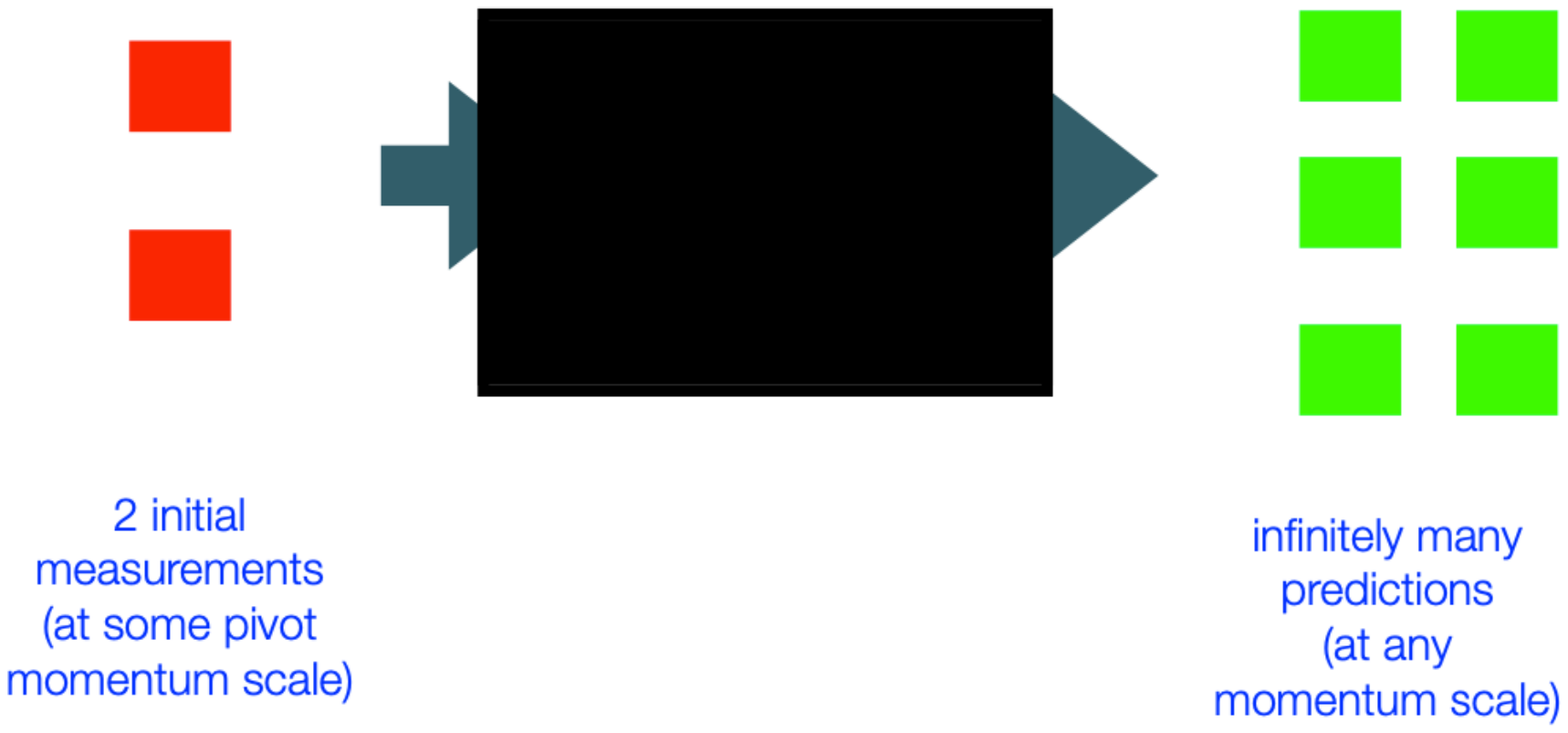}}
\caption{In $\lambda \phi^4$ theory, two initial, finite ``calibrating” measurements enter a black box and return finite predictions for any new experiment.
} \label{BB}
\end{figure}

\begin{figure}[h]
\centerline{\includegraphics[width=1\textwidth, angle = 0]{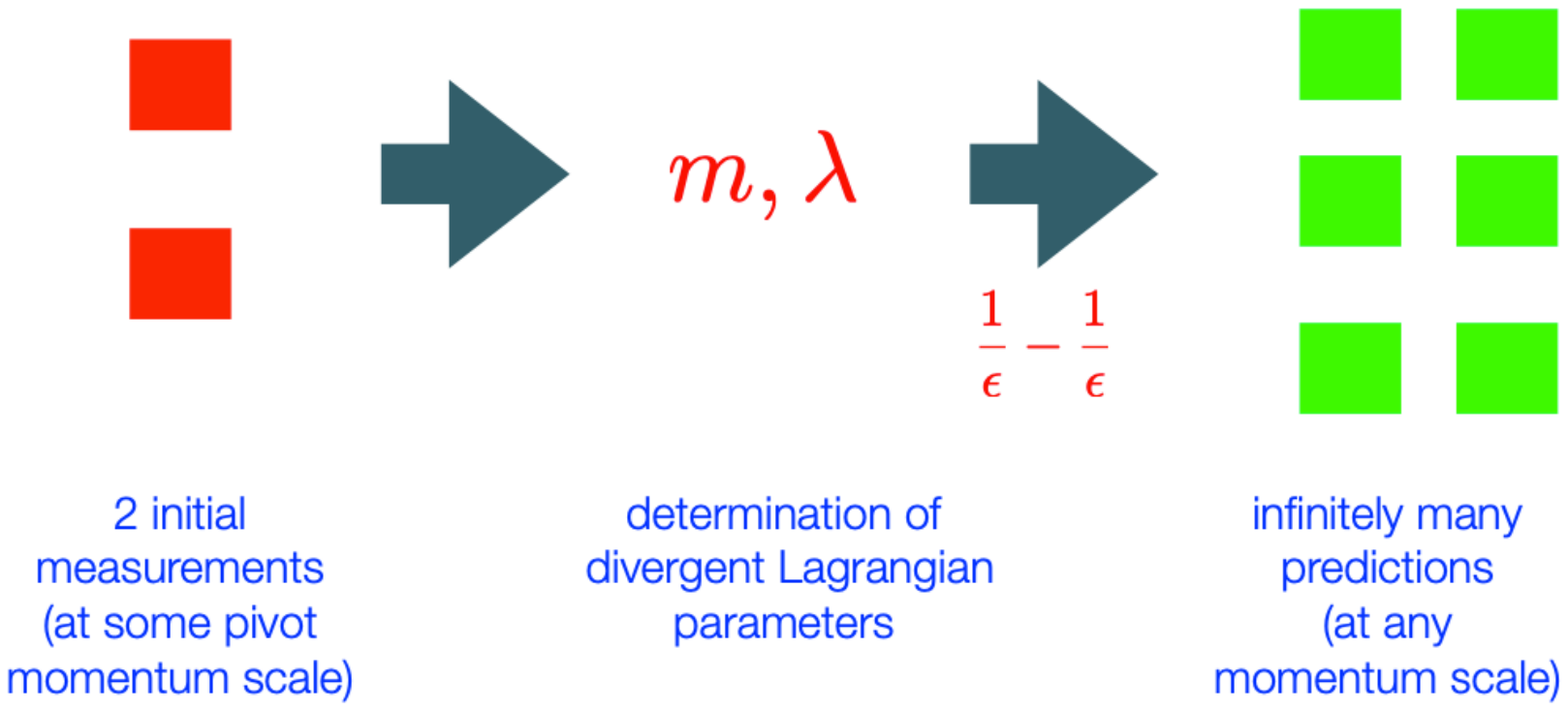}}
\caption{
The textbook approach to renormalisation of $\lambda \phi^4$ theory. Inside the black box remain bare, divergent n-point correlation functions,  computed order by order from Feynman diagrams.  Feeding the same two initial, finite measurements to these n-point correlation functions produce infinite values for the Lagrangian parameters $\lambda$ and $m$. By renormalising the n-point functions, one subtracts one infinity from the other to finally arrive at the same, finite predictions.
} \label{tradiren}
\end{figure}

\begin{figure}[h]
\centerline{\includegraphics[width=1\textwidth, angle = 0]{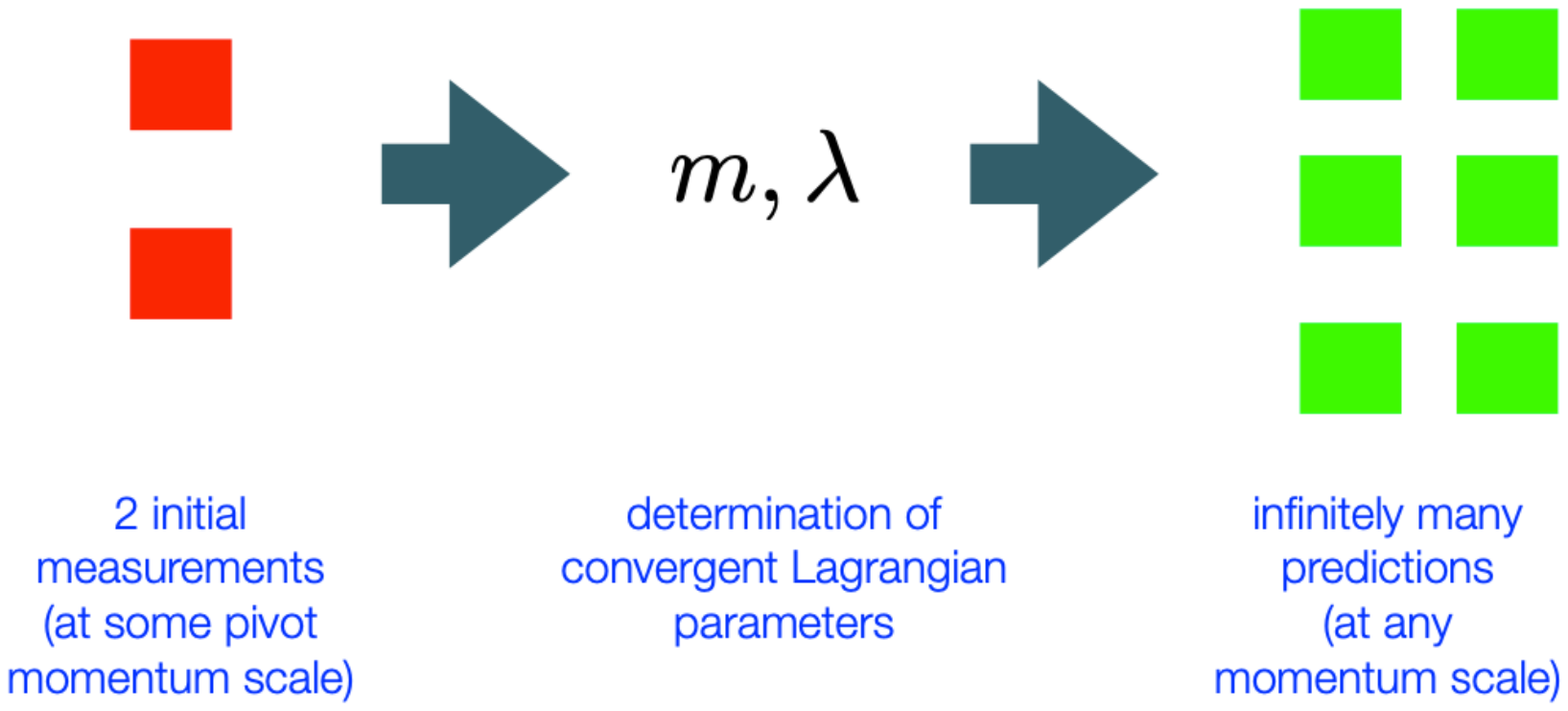}}
\caption{
Proposal for finite renormalisation of $\lambda \phi^4$ theory. By circumventing divergent correlation functions, the same finite initial measurements lead to new, finite expressions for the Lagrangian parameters. No divergent subtractions are needed to produce the same, finite predictions.
} \label{finiren}
\end{figure}

The multiplicative renormalisation procedure is thus an ingenious way to provide the connection between the parameters characterising the theory and experimentally mea\-sura\-ble quantities, such as cross-sections of different reactions. The method respects fundamental requirements to the theory, such as unitarity, causality and Lorentz-invariance.  In its setup, it uses the UV divergent integrals, but at the end of the day, this is just a mapping between the well-defined set of {\em finite}  parameters that characterise the theory and the set of experimental observables. See Figs.~\ref{BB} and \ref{tradiren}. After all, in all physically meaningful theories, finite experimental results must lead to finite predictions for new experiments. It is clear, therefore, that there must exist procedure(s) providing this mapping {\em without encounter of any divergent expressions at any stage of the computation}. See Fig.~\ref{finiren}.

Since (quadratic)  infinities are the core of the hierarchy problem, the question arises whether the fine-tunings (cancellation between large numbers appearing in the traditional formulation of QFT) can be evaded as well.  The existence of such a formalism without large cancellations would challenge the ``naturalness'' paradigm. Indeed, if all expressions are finite, the computation of low-energy observables should not require knowledge of the UV domain of the theory. Moreover, if {\em just one} particular formalism of computations in QFT without the necessity of fine-tunings is found, it will provide a strong argument that the problem of quantum stability of the electroweak scale against radiative corrections is formalism dependent and thus unphysical.

The mere existence of the formulation of QFT without infinities is pretty obvious from the ``mapping'' argument.  And, indeed, such divergence-free methods have been invented already in the past. They are almost not in use in contemporary practical computations in perturbative QFT but are more suitable for a discussion of conceptual issues such as fine-tunings and naturalness. We will overview these approaches below.

The most famous one is that of  Bogolubov-Parasuk-Hepp-Zimmermann (BPHZ) (for the textbook discussion, see \cite{Bogoliubov}). A certain procedure (called ``R-operation'') is applied to any Feynman graph ({\em before} performing integrations over internal momenta) changing the {\em integrand} prescribed by the Feynman rules to another one. The resulting expression is then integrated, with no infinities encountered. The R-operation can be used in both renormalisable and non-renormalisable field theories, as was stressed recently by D. Kazakov \cite{Kazakov:2020mfp}. A refined method, with the use of asymptotics of Feynman {\em integrand}, was proposed recently by F. Tkachov and collaborators \cite{Lenshina:2020edt}. This method is purely perturbative and is applicable to any single Feynman diagram. It does not require any regularisation or renormalisation. Even if the BPHZ approach is well-known, it seems that it has only rarely been explicitly appreciated as a ``convergent counter-example" to the common lore that UV divergences are inescapable.

Yet another scheme is the one based on Callan-Symanzik (CS) equations \cite{Callan:1970yg,Symanzik:1970rt}. Usually, these equations are represented as a tool for the renormalisation group investigation of the high energy behaviour of the renormalised amplitudes. However, it was shown in \cite{Blaer:1974foy,Callan:1975vs} that in fact, the same equations can be used for the construction of the divergence-free and thus completely finite perturbation theory. In this method, the CS equations should be fed by finite perturbative expressions derived from Feynman graphs, and can therefore be called semi-perturbative. 

More recently, 't Hooft \cite{tHooft:2004bkn} proposed yet another method of divergence-free computations in renormalisable quantum field theories.  It consists of exact equations for irreducible two-, three -, and four-point vertices which do not contain any ultraviolet infinities. The idea is that any divergent $n$-point function can be rendered finite by subtracting the same $n$-point function evaluated at different values for the external momenta. This difference can be interpreted as a new irreducible Feynman diagram with $n+1$ external lines. Integrating these ``difference diagrams" with respect to the external momenta yields renormalisation group equations. The parameters of the theory (such as the electron charge and mass in quantum electrodynamics) only enter in terms of freely adjustable integration constants in these equations. Potentially, these equations may result in a completely non-perturbative and divergence-free definition of the theory.  Again, this program avoids all intermediate UV infinities. It seems that, surprisingly, the paper \cite{tHooft:2004bkn} has been largely overlooked in the literature.

In addition to these well-developed methods we can also mention other approaches formulated in the article by Lehmann, Symanzik and Zimmermann \cite{Lehmann:1954rq} (besides the famous LSZ reduction formula it contains the discussion of the finite formulation of QFT), and the work by Nishijima  \cite{Nishijima:1960zz}. Here one attempts to compute renormalised $n$-point functions without even writing down a Lagrangian. Although very tantalising, apparently this approach never got to a mature state.
Other, more algebraic approaches that avoid renormalisation at all can be found in \cite{Kasia,Morgan}. A UV finite QFT with nonlocal field operators, and a subsequent proposal for UV complete quantum gravity, were constructed and further explored in \cite{moff1,moff2,moff3,moff4,moff5}.

To analyse the fine-tunings and the naturalness paradigm, in this work we pick up one of the divergence-free methods mentioned above, based on the Callan-Symanzik equations  (to which we will refer from now on as the ``CS method"). Besides its relative simplicity, it is well adapted to the study of these conceptual problems. We show that the fine-tuning problem indeed does not show up in a simple QFT with two scalar fields with widely separated masses. Although the theory we consider is just a toy model, it incorporates all essential features of the technical hierarchy problem and thus is generic.

On the technical side, our paper does not contain much new  (except for the generalisations of the CS method to more than one field, to the zero-point function and the effective potential). However, the proposal to view the CS method as a finite formulation of QFT (rather than as a proof for renormalizability), and to use that formulation for the discussion of the technical part of the hierarchy problem is new to the best of our knowledge.

The paper is organised as follows. In Section \ref{standard} we provide an overview of the computational methods in QFT, in Section \ref{Cal} we discuss the divergence-free CS method on the example of a single scalar field theory and a two-field scalar theory with hierarchical mass scales. In Section \ref{cc} we discuss the finite QFT approach to the cosmological constant and the effective potential, and we conclude in Section \ref{concl}. Section \ref{outlook}  is an outlook.

\section{Standard approach: multiplicative renormalisation 
 \label{standard}}
 
 One of the goals of QFT is to compute $n$-point Green's functions, which through the LSZ formalism are directly related to physical observables like particle lifetimes and cross-sections. In this section, we review the standard approach to the renormalisation of these functions. This is all textbook material, presented here just to fix the notations,  to remind how the necessity of fine-tunings shows up in the standard approach, and to contrast it later with the finite QFT formulation.

\subsection{Fine-tuning of ``infinite" contributions \label{1fmult}}

Let us consider the simple theory of a single scalar field with a quartic self-interaction:
\be
\L = - \frac{1}{2} \left(\d_\mu \phi\right)  \left(\d^\mu \phi\right) -\frac{m^2}{2}\phi^2 - \frac{\lambda}{4!}\phi^4. \label{lor}
\ee
At the tree level, the contributions to the two- and four-point correlation functions of the field $\phi$ are simply
\ba
\Gamma^{(2)}_{\rm tree} &=& i\left(k^2+m^2\right)~,\nn\\
\Gamma^{(4)}_{\rm tree}&=&-i\lambda~. \label{treel}
\ea

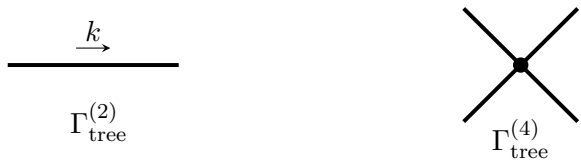
\begin{figure}[!h]
\begin{center}
\begin{tikzpicture}
[line width=1.5 pt, scale=1.5]

\begin{scope}
[shift={(0,0)}]
\draw (-0.5,0)--(1,0);
\node at (0.3,-0.5) {$ \Gamma^{(2)}_{\rm tree}$};
\draw [thin][-stealth](0.1,0.15) -- (0.4,0.15);
\node at (0.25,0.3){$k$};
\end{scope}

\begin{scope}
[shift={(4,0)}]
\draw (-0.5,0.5)--(0.5,-0.5);
\draw (-0.5,-0.5)--(0.5,0.5);
\draw [fill=black](0,0) circle (0.05cm);
\node at (0,-0.65) {$\Gamma^{(4)}_{\rm tree}$};
\end{scope}
\end{tikzpicture}
\caption{ Tree level contributions to the two- and four-point function.
}
\end{center}
\end{figure}

From the one-loop level on, however, the two- and four-point functions receive  contributions that become infinitely large when the internal loop momentum $l$ goes to infinity:
\ba
\Gamma^{(2)} &=& i\left(k^2+m^2\right) +\frac{\lambda}{2}\int \frac{d^4l}{(2\pi)^4}~\frac{1}{l^2+m^2}+\O\left(\lambda^2\right)~,\nn\\
\Gamma^{(4)}&=&-i\lambda+\sum_{\rm 3~opt}\frac{\lambda^2}{2}\int \frac{d^4l}{(2\pi)^4}~\frac{1}{l^2+m^2}~\frac{1}{(l+\kappa_1)^2+m^2}+\O\left(\lambda^3\right)~.
\ea
(Here the sum is over the three Mandelstam channels, i.e. the three distinct ways to pair up the four external momenta. Explicitly, it goes over $\kappa_1 =\{k_1+k_2,k_1+k_3,k_1+k_4\}$.)

\begin{figure}[!h]
\begin{center}
\begin{tikzpicture}
[line width=1.5 pt, scale=1.5]

\begin{scope}
[shift={(0,-0.25)}]
\draw (-0.25,0)--(1.25,0);
\draw [thin][-stealth](-0.15,-0.15) -- (0.15,-0.15);
\node at (0,-0.3) {$k$};
\draw [fill=black](0.5,0) circle (0.05cm);
\draw (0.5,0.25) circle (0.25cm);
\draw [thin][-stealth](0.35,0.6) -- (0.65,0.6);
\node at (0.5,0.75) {$l$};
\node at (0.5,-0.75) {$ \left[-\Gamma^{(2)}\right]_\lambda$};
\end{scope}

\begin{scope}
[shift={(4,0)}]
\draw (0,0.25)--(0.25,0);
\draw (0,-0.25)--(0.25,0);
\draw [fill=black](0.25,0) circle (0.05cm);
\draw (0.5,0) circle (0.25cm);
\draw [fill=black](0.75,0) circle (0.05cm);
\draw (0.75,0)--(1,0.25);
\draw (0.75,0)--(1,-0.25);
\draw [thin][-stealth](0.35,0.35) -- (0.65,0.35);
\draw [thin][stealth-](0.35,-0.35) -- (0.65,-0.35);
\node at (0.5,0.5) {$l$};
\node at (0.5,-0.5) {$\kappa_1+l$};
\draw [thin][-stealth](-0.7,0) -- (-0.4,0);
\node at (-0.55,-0.15){$\kappa_1$};
\node at (0.5,-1) {$ \left[\Gamma^{(4)}\right]_{\lambda^2}$};
\node at (-0.25,0) {$\Big{\{} $};
\end{scope}
\end{tikzpicture}

\caption{One loop contributions to the two- and four-point function. $\left[\Gamma^{(n)}\right]_{\lambda^r}$ denotes the order $\lambda^r$ contribution to $\Gamma^{(n)}$. $\kappa_1$ denotes the sum over the momenta coming in over the two left external lines. The minus sign in front of $\Gamma^{(2)}$ is to remind that in our $(-+++)$ signature, the two-point functions is given by $i\left(k^2+m^2\right)$ minus the sum of all diagrams.
}
\end{center}
\end{figure}
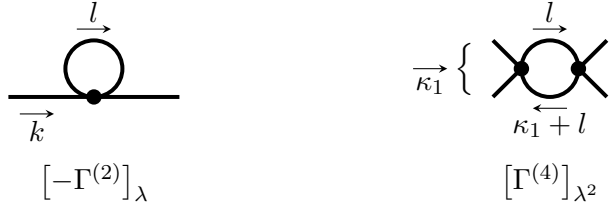

Clearly, something has gone wrong. Since correlation functions are directly related to observables, they must be finite. On top of that, an expansion (in the parameter $\lambda$) with infinite coefficients is meaningless.

In the standard approach to solve this problem, one first regularises the infinities. Here we choose to work with cut-off regularisation: after moving to Euclidean four-momenta the loop integrals over $l$ simply get cut off at some formal value $l_E^2=\Lambda^2$. This gives 
\ba
\Gamma^{(2)} &=& i\left(k^2+m^2\right) +\frac{i\lambda}{32\pi^2}\left(\Lambda^2-m^2\ln{\left(\frac{\Lambda^2}{m^2}\right)}\right)+\O\left(\lambda^2\right)~,\nn\\
\Gamma^{(4)}&=&-i\lambda+\sum_{\rm 3~opt}\frac{i\lambda^2}{32\pi^2}\int_0^1 dx~\ln{\left(\frac{\Lambda^2}{x(1-x)\kappa_1^2+m^2}\right)}+\O\left(\lambda^3\right)~.
\ea
Here $x$ is a Feynman parameter. In the limit $\Lambda \to \infty$, we formally recover the original UV divergent integrals that we began with.

Having regularised the UV divergent integrals, we move to the next step: renormalisation. Here one actually subtracts the divergences. This is formalised by adding so-called counterterms to the Lagrangian. We now work with
\ba
\L& =& - \frac{1}{2} \left(\d_\mu \phi\right)  \left(\d^\mu \phi\right) - \frac{m^2}{2} \phi^2 - \frac{\lambda}{4!} \phi^4\nn\\
&& \qquad \qquad \qquad - \frac{1}{2} ~\delta Z \left(\d_\mu \phi\right) \left(\d^\mu \phi\right) - \frac{\delta m^2}{2} \phi^2 - \frac{\delta\lambda}{4!} \phi^4 \ .
\label{Lren}
\ea
The step from the original Lagrangian in eq.~(\ref{lor}) to this new Lagrangian is commonly described as replacing  ``bare" quantities by ``renormalised" quantities. Here we just omit these formal steps. Whatever formal explanation one chooses, effectively the standard approach to renormalisation just comes down to adding the counterterms $\delta Z$, $\delta m^2$ and $\delta \lambda$ to the original Lagrangian. Now the two-and four-point functions read
\ba
\Gamma^{(2)} &=& i\left(k^2+m^2\right) +\frac{i\lambda}{32\pi^2}\left(\Lambda^2-m^2\ln{\left(\frac{\Lambda^2}{m^2}\right)}\right)+i\delta Z\cdot k^2+i\delta m^2+\O\left(\lambda^2\right)~,\nn\\
\Gamma^{(4)}&=&-i\lambda+\sum_{\rm 3~opt}\frac{i\lambda^2}{32\pi^2}\int_0^1 dx~\ln{\left(\frac{\Lambda^2}{x(1-x)\kappa_1^2+m^2}\right)}-i \delta \lambda+\O\left(\lambda^3\right)~.
\ea
Next the counterterms $\delta Z$, $\delta m^2$ and $\delta \lambda$ can be chosen such that they ``absorb" the UV divergences. In fact, there are infinitely many ways to do so, since we can absorb any (local) finite contribution as well. Here we choose the so-called zero external momentum subtraction renormalisation scheme. This scheme is designed in such a way that at $k^2=0$, all quantum corrections vanish. It is realised by picking
\ba
\delta Z &=& \O\left(\lambda^2\right)~,\nn\\
\delta m^2 &=& -\frac{\lambda}{32\pi^2}\left(\Lambda^2-m^2\ln{\left(\frac{\Lambda^2}{m^2}\right)}\right)+ \O\left(\lambda^2\right)~,\nn\\
\delta \lambda &=& \frac{3\lambda^2}{32\pi^2}\ln{\left(\frac{\Lambda^2}{m^2}\right)}+\O\left(\lambda^3\right)~.
\ea
We end up with finite, renormalised two- and four-point functions (the overbar on $\Gamma$ denotes that these are \emph{renormalised} Green's functions)
\ba
\bar{\Gamma}^{(2)} &=& i\left(k^2+m^2\right) +\O\left(\lambda^2\right)~,\nn\\
\bar{\Gamma}^{(4)}&=&-i\lambda+\sum_{\rm 3~opt}\frac{i\lambda^2}{32\pi^2}\int_0^1 dx~\ln{\left(\frac{m^2}{x(1-x)\kappa_1^2+m^2}\right)}+\O\left(\lambda^3\right)~. \label{gamren}
\ea
UV divergences coming from higher loop diagrams can be absorbed in higher-order contributions to the counterterms. We also note a peculiarity of the chosen renormalisation scheme: the order $\lambda$ contribution to $\bar{\Gamma}^{(2)}$ vanishes. This is caused by the simple fact that in the corresponding one-loop diagram, there is no dependence on the external momentum $k$. Demanding quantum corrections to vanish at some arbitrarily chosen scale (in this case: $k^2=0$) then implies that they vanish at all values for $k^2$.

\begin{figure}[h]
\centerline{\includegraphics[width=1\textwidth, angle = 0]{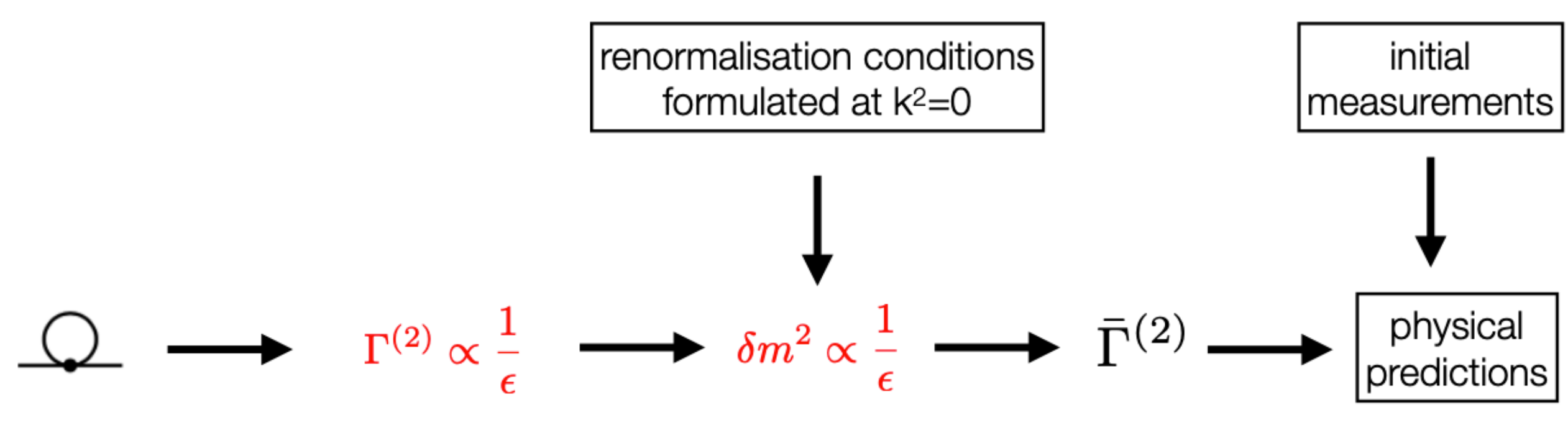}}
\caption{Textbook computation of the renormalised two-point  function $\bar{\Gamma}^{(2)}$ in $\lambda \phi^4$ theory, using a mass-dependent renormalisation scheme. Feynman graph integration of loop diagrams yields divergent contributions to bare n-point functions. Renormalisation conditions (here defined at $k^2=0$) prescribe the divergent subtraction (formalised in counterterms) needed to obtain the renormalised two-point function $\bar{\Gamma}^{(2)}$. Combining the n-point functions with some initial measurements, scheme ambiguities disappear and predictions for physical observables like cross sections and decay rates can be made.} \label{figtrad}
\end{figure}

Although this standard approach to renormalisation, summarised in Fig. \ref{figtrad}, works perfectly fine, it leaves the uncomfortable impression that formally infinite fine-tunings between loop corrections and counterterms are necessary to obtain finite results for the renormalised correlation functions. In the words of Collins  \cite{Collins:1984xc}, describing the situation with multiplicative renormalisation: ``The presence of ultra-violet divergences, even though they are cancelled by renormalisation counterterms, means that in any process there are contributions from quantum fluctuations on every distance scale."

\subsection{Fine-tuning of ``large" contributions}

The problem described in the previous section becomes more acute once we consider models with concrete realisations of ``UV physics". Let us consider a model of two interacting scalar fields
\be
\L = - \frac{1}{2} \left(\d_\mu \phi\right)  \left(\d^\mu \phi\right)- \frac{1}{2} \left(\d_\mu \Phi\right)  \left(\d^\mu \Phi\right) - \frac{m^2}{2} \phi^2- \frac{M^2}{2} \Phi^2 - \frac{\lambda_\phi}{4!} \phi^4 - \frac{\lambda_{\phi\Phi}}{4} \phi^2\Phi^2 - \frac{\lambda_{\Phi}}{4!} \Phi^4~. \label{L2}
\ee
We assume that $m \ll M$. The physics involving the field $\Phi$, therefore, acts as a toy representation of ``new physics" living at large energy scales. We assume there is no other physics in between these scales: we are in the rightmost situation in Figure \ref{plaatjemisha}. To make mea\-ning\-ful perturbative expansions, we need to assume that the three four-point couplings $\lambda_i = \{\lambda_\phi,\lambda_{\phi\Phi}, \lambda_\Phi\}$ are of the same order. This order will be generally denoted as $\lambda$. This theory represents an ideal playground for the analysis of fine-tuning problems. The low energy effective theory of the light scalar field is given by (\ref{lor})  plus higher-dimensional operators suppressed by the large mass $M^2$. The exact form of these operators can be found by integrating out of the heavy field.

For our purposes, it is sufficient to consider the two-point function of the light field $\phi$. We will refer to it as $\Gamma^{(2\phi)}$. At the one-loop level, it receives quantum corrections from a diagram with a ``light" $\phi$ loop and another diagram with a ``heavy" $\Phi$ loop.

 \begin{figure}[!h]
\begin{center}
\begin{tikzpicture}
[line width=1.5 pt, scale=1.5]

\begin{scope}
\node at (0,0) {$\Gamma^{(2\phi)}$};
\end{scope}

\begin{scope}
\node at (0.5,0) {$=$};
\end{scope}

\begin{scope}
[shift={(1,0)}]
\draw (0,0)--(0.75,0);
\node at (1.125,0) {$-$};
\end{scope}

\begin{scope}
[shift={(2.5,0)}]
\draw (0,0)--(1,0);
\draw [fill=black](0.5,0) circle (0.05cm);
\draw (0.5,0.25) circle (0.25cm);
\node at (1.25,0) {$-$};
\end{scope}

\begin{scope}
[shift={(4,0)}]
\draw (0,0)--(1,0);
\draw [fill=black](0.5,0) circle (0.05cm);
\draw [dashed] (0.5,0.25) circle (0.25cm);
\node at (1.75,0) {$+~~\O\left(\lambda^2\right)$};
\end{scope}
\end{tikzpicture}

\end{center}

\caption{Tree level and one-loop contribution to the light field $\phi$'s two-point function. $\phi$ is denoted by continuous lines, and $\Phi$ is denoted by dashed lines.}

\end{figure}

Again, we first need to regularise the UV infinities coming from the loop correction diagrams. Now we choose to work with dimensional regularisation, which is not sensitive to power-like divergences\footnote{In the single field case considered in the previous sub-section there is no fine-tuning problem in the MSbar renormalisation scheme, just because there is only one mass scale present in the problem.}. We evaluate the Feynman integrals in $d=4-2\eps$. Now UV divergences get isolated in $1/\eps$ poles. Applying the MSbar renormalisation scheme, i.e. subtracting the UV divergent terms through a suitable choice of counterterms, leaves us with
\be
\bar{\Gamma}^{(2\phi)} = i(k^2+m^2)-\frac{i \lambda_{\phi} m^2}{32\pi^2}\left( 1 +\ln \frac{ \mu^2}{m^2} \right) -\frac{i \lambda_{\phi\Phi} M^2}{32\pi^2}\left( 1 +\ln \frac{ \mu^2}{M^2} \right)~.\label{msb}
\ee

We now see that even after subtracting the formal UV divergences proportional to $1/\eps$,  $\bar{\Gamma}^{(2\phi)}$ still receives large contributions of order $M^2$.  Therefore, it seems that the heavy-scale physics of order $M^2$ has a dramatic influence on the physics of order $m^2$. In the context of the Standard Model, explaining the small Higgs mass that we observe requires a tremendous fine-tuning between the bare Higgs mass $m^2$ and its one-loop corrections proportional to the scale of new physics $M^2$. This is exactly the essence of the so-called problem of stability of the Higgs mass against radiative corrections which appears in the multiplicative approach to QFT.

In the next Section, we show that no fine-tunings are necessary for the divergence-free approach to QFT.

\section{The Callan-Symanzik method as a finite approach to QFT  
\label{Cal}}

The multiplicative renormalisation procedure provides a recipe for how to convert the limited number of finite parameters characterising the renormalisable theory to specific predictions via the regularisation of infinite expressions and fine-tuning of counterterms to remove the infinities. As we have already discussed in the Introduction, the mapping of the set of finite numbers into another set of finite numbers can be done without any intermediate infinities. The corresponding approach can be called the divergence-less QFT formulation or finite QFT.  In this Section, we explain how a finite QFT -- based on equations reminiscent of the Callan-Symanzik equations -- is constructed, and provide explicit rules showing how it works. Our discussion treats the one-loop case in detail and sketches the generalisation to higher loop orders. 
We will refer to the method as the ``CS method".  We will take the examples considered in the previous Section.

This alternative approach to renormalisation in QFT was proposed by Blaer and Young \cite{Blaer:1974foy}, and Callan  \cite{Callan:1970yg} in his lecture notes for the Les Houches summer school in 1975. The quote from these notes: ``The method we shall use to prove renormalisability is not, strictly speaking, new but is certainly not generically known"  looks to be valid even today. In this remarkable program, all quantum corrections to all $n$-point functions can be found, order by order in the coupling constant, from only finite ingredients:
\begin{enumerate}
\item  Manifestly convergent (connected) diagrams (diagrams with the degree of divergence less than zero).
\item  A set of equations (reminiscent of the well-known Callan-Symanzik equations) between $n$-point functions and their derivatives with respect to the mass parameter.
\item  Boundary conditions.
\end{enumerate}

Although the Callan-Symanzik method was designed to prove the validity of the standard multiplicative renormalisation program, it constitutes by itself an independent approach to the computations in QFT.  Now we take it as a {\bf formulation} of finite Quantum Field Theory. 

\subsection{Divergenceless QFT of one scalar field\label{CS1}}
We begin again from the Lagrangian 
\be
\L = - \frac{1}{2} \left(\d_\mu \phi\right)  \left(\d^\mu \phi\right) -\frac{m^2}{2}\phi^2 - \frac{\lambda}{4!}\phi^4~.
\label{phi4}
\ee
In this Lagrangian, the parameters $m^2$ and $\lambda$ are now {\em finite} numbers. All physical quantities (such as the mass of the scalar particle and different cross-sections) will be eventually expressed through them.  We would like to find $\phi$'s two- and four-point functions in a way that no divergences show up at any stage of the computation.

The tree,  lowest order contributions to these $n$-point functions are found from eq. (\ref{phi4}) in the usual way and are equal to
\ba
\left[\bar{\Gamma}^{(2)}\right]_{\lambda^0} &=& i\left(k^2+m^2\right)~,\nn\\
\left[\bar{\Gamma}^{(4)}\right]_\lambda&=&-i\lambda~. \label{bountree}
\ea
Of course, no divergences are met at this stage.

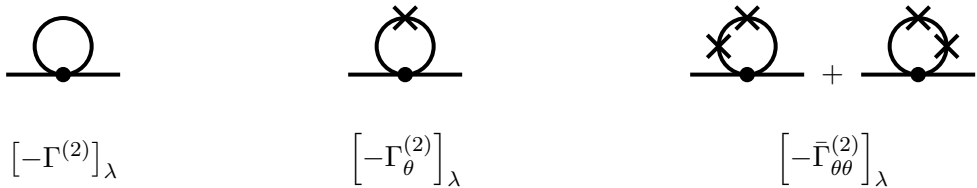
\begin{figure}[!h]
\begin{center}
\begin{tikzpicture}
[line width=1.5 pt, scale=1.5]

\begin{scope}
\draw (0,0)--(1,0);
\draw [fill=black](0.5,0) circle (0.05cm);
\draw (0.5,0.25) circle (0.25cm);
\node at (0.5,-0.75) {$ \left[-\Gamma^{(2)}\right]_\lambda$};
\end{scope}

\begin{scope}
[shift={(3,0)}]
\draw (0,0)--(1,0);
\draw [fill=black](0.5,0) circle (0.05cm);
\draw (0.5,0.25) circle (0.25cm);
\draw (0.4,0.4)--(0.6,0.6);
\draw (0.4,0.6)--(0.6,0.4);
\node at (0.5,-0.75) {$ \left[-\Gamma^{(2)}_\theta\right]_\lambda$};
\end{scope}

\begin{scope}
[shift={(6,0)}]
\draw (0,0)--(1,0);
\draw [fill=black](0.5,0) circle (0.05cm);
\draw (0.5,0.25) circle (0.25cm);
\draw (0.4,0.4)--(0.6,0.6);
\draw (0.4,0.6)--(0.6,0.4);
\draw (0.15,0.15)--(0.35,0.35);
\draw (0.35,0.15)--(0.15,0.35);
\node at (1.25,-0.75) {$ \left[-\bar{\Gamma}^{(2)}_{\theta\theta}\right]_\lambda$};
\node at (1.25,0) {$+$};
\end{scope}

\begin{scope}
[shift={(7.5,0)}]
\draw (0,0)--(1,0);
\draw [fill=black](0.5,0) circle (0.05cm);
\draw (0.5,0.25) circle (0.25cm);
\draw (0.4,0.4)--(0.6,0.6);
\draw (0.4,0.6)--(0.6,0.4);
\draw (0.65,0.15)--(0.85,0.35);
\draw (0.65,0.35)--(0.85,0.15);
\end{scope}

\end{tikzpicture}

\caption{One and two $\theta$-operations on  the one loop (order $\lambda$) correction to $\phi$'s two-point function.  \label{thet2}}

\end{center}
\end{figure}

\begin{figure}[!h]
\begin{center}
\begin{tikzpicture}
[line width=1.5 pt, scale=1.5]

\begin{scope}
\draw (0,0.25)--(0.25,0);
\draw (0,-0.25)--(0.25,0);
\draw [fill=black](0.25,0) circle (0.05cm);
\draw (0.5,0) circle (0.25cm);
\draw [fill=black](0.75,0) circle (0.05cm);
\draw (0.75,0)--(1,0.25);
\draw (0.75,0)--(1,-0.25);
\node at (0.5,-0.75) {$ \left[\Gamma^{(4)}\right]_{\lambda^2}$};
\end{scope}

\begin{scope}
[shift={(3,0)}]
\draw (0,0.25)--(0.25,0);
\draw (0,-0.25)--(0.25,0);
\draw [fill=black](0.25,0) circle (0.05cm);
\draw (0.5,0) circle (0.25cm);
\draw (0.4,0.15)--(0.6,0.35);
\draw (0.4,0.35)--(0.6,0.15);
\draw [fill=black](0.75,0) circle (0.05cm);
\draw (0.75,0)--(1,0.25);
\draw (0.75,0)--(1,-0.25);
\node at (1.25,-0.75) {$ \left[\bar{\Gamma}^{(4)}_\theta\right]_{\lambda^2}$};
\node at (1.25,0) {$+$};
\end{scope}

\begin{scope}
[shift={(4.5,0)}]
\draw (0,0.25)--(0.25,0);
\draw (0,-0.25)--(0.25,0);
\draw [fill=black](0.25,0) circle (0.05cm);
\draw (0.5,0) circle (0.25cm);
\draw (0.4,-0.35)--(0.6,-0.15);
\draw (0.4,-0.15)--(0.6,-0.35);
\draw [fill=black](0.75,0) circle (0.05cm);
\draw (0.75,0)--(1,0.25);
\draw (0.75,0)--(1,-0.25);
\end{scope}

\end{tikzpicture}

\caption{$\theta$-operation on  the one loop (order $\lambda^2$) correction to $\phi$'s four-point function.  \label{thet4}}

\end{center}
\end{figure}
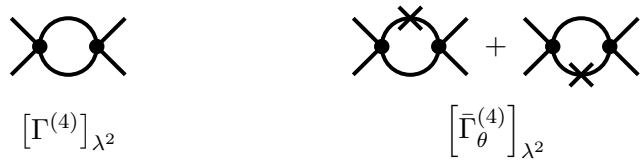

To find the quantum corrections to these correlation functions, we need to define a so-called ``$\theta$-operation." In its graphical representation\footnote{There exists another, algebraic, representation of the $\theta$-operation which we will discuss in our companion paper \cite{SanderMisha1}. See also the original works \cite{Blaer:1974foy,Callan:1975vs}.}, it can be applied to Feynman diagrams. This o\-pe\-ra\-tion splits every propagator, one by one, in two parts by inserting a new kind of ``cross" vertex, which comes with Feynman rule $(-1)$. Effectively, applying the $\theta$-operation on a diagram with $n$ propagators yields $n$ new diagrams which each have $(n+1)$ propagators. All new ``theta-fied" diagrams are copies of the original diagram, but with one particular propagator cut in two. See figures \ref{thet2} and \ref{thet4}.

Two short (and manifestly finite) computations yield 
\ba
\left[\bar{\Gamma}^{(2)}_{\theta \theta}\right]_\lambda &=& -\frac{i\lambda}{32\pi^2}~\frac{1}{m^2}~,\nn\\
\left[\bar{\Gamma}^{(4)}_\theta\right]_{\lambda^2} &=&  -\frac{\lambda^2}{32\pi^2}\sum_{\rm 3~opt}~\int_0^1 dx~ \frac{1} {x(1-x)\kappa_1^2+m^2}~.\label{gtet}
\ea
These diagrams are the diagrams that are mentioned in point (i) on page 10.

Next, these results are fed into the following equations (point (ii) on page 10), for a derivation, see the original works \cite{Blaer:1974foy,Callan:1975vs}, and our companion paper \cite{SanderMisha1}) which are now considered to be a part of the definition of the finite QFT and thus simply postulated
\ba
2i m^2 ~(1+\gamma)\cdot\bar{\Gamma}^{(n)}_\theta&=& \left[  \left(m ~\frac{\d}{\d m} +\beta~\frac{\d}{\d \lambda}     \right)+n\cdot \gamma\right] \bar{\Gamma}^{(n)}\nn~,\\
2im^2~(1+\gamma)\cdot  \bar{\Gamma}^{(n)}_{\theta \theta} &=& \left(m~\frac{\d}{\d m} +\beta~\frac{\d}{\d \lambda}+n\cdot\gamma+\gamma_\theta\right)\bar{\Gamma}^{(n)}_\theta~.
\label{calrel}
\ea
Here $\beta$, $\gamma$ and $\gamma_\theta$ are additional {\em finite} quantities that follow, order by order, in the computation of the $n$-point functions that we are after. It happens that $\beta$ corresponds to the standard $\beta-$function for the scalar self-coupling, and $\gamma$ is related to the field renormalisation. For now, we just need to postulate that they are independent of external momenta and that $\beta$ begins at order $\lambda^2$, while both $\gamma$ and $\gamma_\theta$ begin at order $\lambda$.

These equations are supplemented by the following boundary conditions\footnote{In the multiplicative renormalisation procedure, these conditions specify the mass-dependent, zero external momentum subtraction renormalisation scheme, see e.g. \cite{Peskin:1995ev}. A generalisation of the CS method that does not require such boundary conditions was introduced in \cite{Naud:1998yg}. \label{foot}} (point (iii) on page 10), valid up to all orders in $\lambda$. These conditions are:
\be
\left[\frac{d}{dk^2}~\bar{\Gamma}^{(2)}(k^2)\right]_{k^2 = 0} =i, \qquad  \bar{\Gamma}^{(2)}\left(k^2 = 0\right) =i m^2,\qquad  \bar{\Gamma}^{(4)}\left(k^2 = 0\right) =-i\lambda~.
\label{bounren}
\ee
We see that the boundary conditions are imposed at $k^2=0$, as was done in the original papers \cite{Blaer:1974foy,Callan:1975vs}. In principle, this method works for any choice for the external momentum scale where the boundary conditions are formulated. The particular choice $k^2=0$ simplifies computations but renders the method ill-defined for massless fields. Another possibility is $k^2=-m^2$, which corresponds to the physical, on-shell renormalisation scheme.

Note that when evaluating the first equation in eq.~(\ref{calrel})  at $k^2=0$, the boundary conditions directly imply 
\be
\bar{\Gamma}^{(2)}_\theta(k^2=0) = 1, \label{gtt2a}
\ee
up to all orders in $\lambda$. Then it looks natural to formulate the last ``tree level" boundary condition that completes the system as
\be
\left[\bar{\Gamma}^{(2)}_\theta\right]_{\lambda^0} = 1. \label{gtt3a}
\ee
To be very clear: eq.~(\ref{gtt2a}) is a result of the system of equations, while eq.~(\ref{gtt3a}) is one of the postulates on which the system is built.

Now let us illustrate how one recovers the $n$-point functions from these rules and equations.

\subsubsection{Four-point function, $n=4$}

 Here it is enough to only consider the first equation in eq.~(\ref{calrel}). Inserting the second boundary condition in eq.~(\ref{bountree}) we see that the first contributions to this equation occur at order $\lambda^2$. There we have

\be  
-\frac{i\lambda^2m^2}{16\pi^2}\sum_{\rm 3~opt}~\int_0^1 dx~ \frac{1} {x(1-x)\kappa_1^2+m^2} = m~\frac{\d}{\d m}\left[\bar{\Gamma}^{(4)}\right]_{\lambda^2} -i (\left[\beta\right]_{\lambda^2} +4\left[\gamma\right]_\lambda \cdot \lambda). \label{receq1f}
\ee
Evaluating this equation at $k^2=0$ we can use the boundary condition $\bar{\Gamma}^{(4)}(0)=-i\lambda$, and find
\be
-\frac{3i\lambda^2}{16\pi^2} = -i\left( \left[\beta\right]_{\lambda^2}+4\lambda  \left[\gamma\right]_\lambda   \right). \label{bgam}
\ee
Subtracting this relation from eq.~(\ref{receq1f}) leaves us with 
\be
 \frac{\d}{\d m^2}\left[\bar{\Gamma}^{(4)}\right]_{\lambda^2}=-\frac{i\lambda^2}{32\pi^2}\sum_{\rm 3~opt}~\int_0^1 dx~ \frac{1} {x(1-x)\kappa_1^2+m^2} +\frac{3i\lambda^2}{32\pi^2}\cdot \frac{1}{m^2}.
 \ee
Integrating and using the same boundary conditions yields the familiar result of eq.~(\ref{gamren}), but now obtained in a manifestly finite way,
\be
\bar{\Gamma}^{(4)}= -i\lambda+\frac{i\lambda^2}{32\pi^2}\sum_{\rm 3~opt}~\int_0^1 dx~ \ln{\frac{m^2}{x(1-x)\kappa_1^2+m^2}}+\O\left(\lambda^3\right).
\ee

\subsubsection{Two point function, $n=2$}

Here, the final goal is to derive $\left[\bar{\Gamma}^{(2)}\right]_\lambda$ through the first relation in eq.~(\ref{calrel}). However, to do so we first need to find $\left[\bar{\Gamma}_\theta^{(2)}\right]_\lambda$ from the second one. Inserting everything that we know, this equation reads at order $\lambda$
\be
\frac{\lambda}{16\pi^2} = m\frac{\d}{\d m}\left[\bar{\Gamma}^{(2)}_\theta\right]_\lambda + 2\left[\gamma\right]_\lambda+\left[\gamma_\theta\right]_\lambda.
\ee
At $k^2=0$ the boundary conditions in eq.~(\ref{bounren}) then give
\be
\frac{\lambda}{16\pi^2} =2\left[\gamma\right]_\lambda+\left[\gamma_\theta\right]_\lambda. \label{bgam2}
\ee
Subtracting this one from the previous equation, and using the boundary conditions again gives
\be
m\frac{\d}{\d m}\left[\bar{\Gamma}^{(2)}_\theta\right]_\lambda =0 \qquad \Rightarrow\qquad \bar{\Gamma}^{(2)}_\theta=1+\O\left(\lambda^2\right).
\ee
With this, we are ready to return to the first relation in eqs.~(\ref{calrel}), and evaluate it for $n=2$. At order $\lambda^0$ it gives just a consistency check: $2im^2=2im^2$. At order $\lambda$, we get
\be
2im^2\left[\gamma\right]_\lambda =
m~\frac{\d}{\d m}\left[\bar{\Gamma}^{(2)}\right]_\lambda +2\left[\gamma\right]_\lambda\cdot i\left(k^2+m^2\right).\label{redeq}
\ee
Inserting the first boundary condition from eq.~(\ref{bounren}) gives just
\be
0=\left[\gamma\right]_\lambda. \label{bgam3}
\ee
This goes back into eq.~(\ref{redeq}):
\be
m~\frac{\d}{\d m}\left[\bar{\Gamma}^{(2)}\right]_\lambda=0,
\ee
so upon using the renormalisation condition $\bar{\Gamma}^{(2)}(0) = i m^2$ up to all orders in $\lambda$ we finally recover the standard result from eq.~(\ref{gamren}) without having made a divergent subtraction
\be
\left[\bar{\Gamma}^{(2)}\right]_\lambda=0 \qquad \leftrightarrow\qquad \bar{\Gamma}^{(2)} =i\left(k^2+m^2\right)+\mathcal{O}\left(\lambda^2\right).
\ee
Of course, the absence of an order $\lambda$ contribution to $\bar{\Gamma}^{(2)}$ ultimately follows from the fact that at order $\lambda$, there is no dependence on the external momenta in $\bar{\Gamma}^{(2)}_{\theta\theta}$.

\subsubsection{Higher loop order}

So far, we have fed $\left[\bar{\Gamma}^{(4)}_\theta\right]_{\lambda^2}$ and $\left[\bar{\Gamma}^{(2)}_{\theta\theta}\right]_\lambda$ to the system of differential equations and the associated boundary conditions. We have obtained the quantities that we were after, $\left[\bar{\Gamma}^{(2)}\right]_{\lambda}$ and $\left[\bar{\Gamma}^{(4)}\right]_{\lambda^2}$. Also, as byproducts we have found the additional quantities $\left[\bar{\Gamma}^{(2)}_\theta\right]_\lambda$ and (from eqs.~(\ref{bgam}, \ref{bgam2}) and (\ref{bgam3})), $\left[\beta\right]_{\lambda^2}$, $\left[\gamma\right]_\lambda$ and $\left[\gamma_\theta\right]_\lambda$ respectively.

What comes next? How do we find $\left[\bar{\Gamma}^{(2)}\right]_{\lambda^2}$ and $\left[\bar{\Gamma}^{(4)}\right]_{\lambda^3}$? The system of equations and boundary conditions are still the same. We just have to feed them $\left[\bar{\Gamma}^{(4)}_\theta\right]_{\lambda^3}$ and $\left[\bar{\Gamma}^{(2)}_{\theta\theta}\right]_{\lambda^2}$. As byproducts, we will find $\left[\bar{\Gamma}^{(2)}_\theta\right]_{\lambda^2}$, $\left[\beta\right]_{\lambda^3}$, $\left[\gamma\right]_{\lambda^2}$ and $\left[\gamma_\theta\right]_{\lambda^2}$.

The problem therefore reduces to obtaining $\left[\bar{\Gamma}^{(4)}_\theta\right]_{\lambda^3}$ and $\left[\bar{\Gamma}^{(2)}_{\theta\theta}\right]_{\lambda^2}$. There exist two kinds of contributions. On the one hand, we can take the diagrams for $\left[\bar{\Gamma}^{(4)}_\theta\right]_{\lambda^2}$ and $\left[\bar{\Gamma}^{(2)}_{\theta\theta}\right]_\lambda$ that we had already, and insert the found one-loop corrections to the two- and four-point functions. We compute the same diagrams, but with one loop quantum corrected propagators and vertices. It is just the usual skeleton expansion in disguise. (And of course, for $n>4$ all connected $n$-point diagrams are convergent by themselves and can be computed in the usual way through the skeleton expansion.)

On the other hand, there are new diagrams to consider. The new contributions to $\bar{\Gamma}^{(2)}_{\theta\theta}$ follow from performing two $\theta$-operations on the two-loop correction diagram to the two-point function (the so-called sunset diagram). The new contributions to $\bar{\Gamma}^{(4)}_{\theta}$ follow from performing one $\theta$-operation on the two-loop correction diagram to the four-point function. See figure \ref{newdeg}.

\begin{figure}[!h]
\begin{center}
\begin{tikzpicture}
[line width=1.5 pt, scale=1.5]

\begin{scope}
\draw (0,0)--(1,0);
\draw [fill=black](0.25,0) circle (0.05cm);
\draw [fill=black](0.75,0) circle (0.05cm);
\draw (0.5,0) circle (0.25cm);
\draw (0.4,0.15)--(0.6,0.35);
\draw (0.4,0.35)--(0.6,0.15);
\draw (0.4,-0.35)--(0.6,-0.15);
\draw (0.4,-0.15)--(0.6,-0.35);
\node at (-1.5,0) {$ -\bar{\Gamma}^{(2)}_{\theta\theta}~~\supset~~$};
\node at (-0.5,0) {$ 6~~\times$};
\node at (1.5,0) {$ +$};
\end{scope}

\begin{scope}
[shift={(2.75,0)}]
\draw (0,0)--(1,0);
\draw [fill=black](0.25,0) circle (0.05cm);
\draw [fill=black](0.75,0) circle (0.05cm);
\draw (0.5-0.1+0.15,0.19-0.1)--(0.5+0.1+0.15,0.19+0.1);
\draw (0.5+0.1+0.15,0.19-0.1)--(0.5-0.1+0.15,0.19+0.1);
\draw (0.5+0.1-0.15,0.19-0.1)--(0.5-0.1-0.15,0.19+0.1);
\draw (0.5-0.1-0.15,0.19-0.1)--(0.5+0.1-0.15,0.19+0.1);
\draw (0.5,0) circle (0.25cm);
\node at (-0.5,0) {$ 6~~\times$};
\end{scope}

\begin{scope}
[shift={(0,-1)}]
\draw (0,0)--(1,0);
\draw [fill=black](0.25,0) circle (0.05cm);
\draw [fill=black](0.75,0) circle (0.05cm);
\draw (0.5,0) circle (0.25cm);
\draw [fill=black](0.5,0.25) circle (0.05cm);
\draw (0.5,0.25)--(0.75,0.5);
\draw (0.5,0.25)--(0.25,0.5);
\draw (0.4,-0.35)--(0.6,-0.15);
\draw (0.4,-0.15)--(0.6,-0.35);
\node at (-1.5,0) {$ \bar{\Gamma}^{(4)}_{\theta}~~\supset~~$};
\node at (-0.5,0) {$ 2~~\times$};
\node at (1.5,0) {$ +$};
\end{scope}

\begin{scope}
[shift={(2.75,-1)}]
\draw (0,0)--(1,0);
\draw [fill=black](0.25,0) circle (0.05cm);
\draw [fill=black](0.75,0) circle (0.05cm);
\draw (0.5,0) circle (0.25cm);
\draw [fill=black](0.5,0.25) circle (0.05cm);
\draw (0.5,0.25)--(0.75,0.5);
\draw (0.5,0.25)--(0.25,0.5);
\draw (0.5-0.1+0.15,0.19-0.1)--(0.5+0.1+0.15,0.19+0.1);
\draw (0.5+0.1+0.15,0.19-0.1)--(0.5-0.1+0.15,0.19+0.1);
\node at (-0.5,0) {$ 2~~\times$};
\end{scope}

\end{tikzpicture}
\caption{Two loop contributions to $\bar{\Gamma}^{(2)}_{\theta\theta}$ and $\bar{\Gamma}^{(4)}_\theta$.\label{newdeg}}
\end{center}
\end{figure}

\begin{figure}[h]
\centerline{\includegraphics[width=1\textwidth, angle = 0]{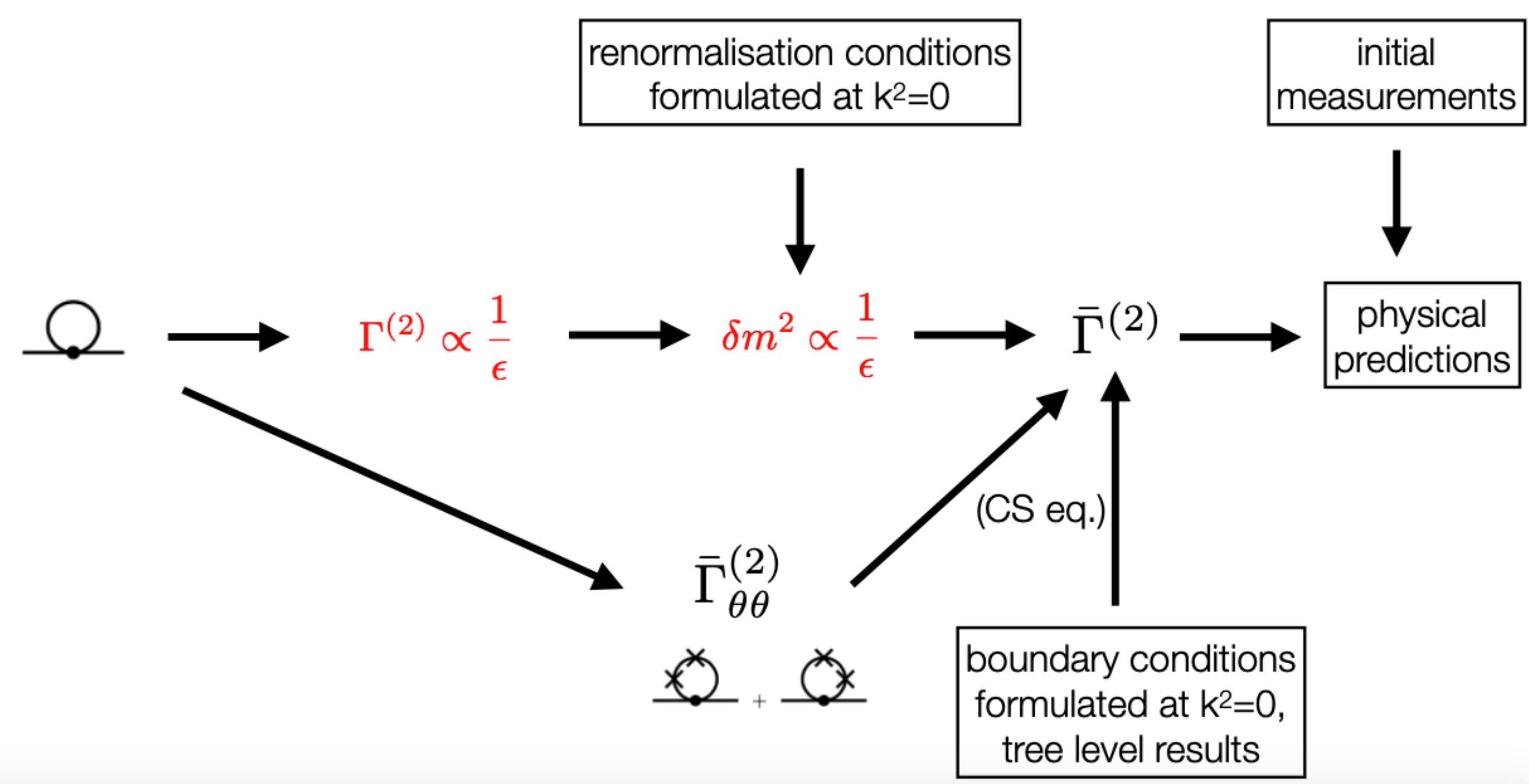}}
\caption{The CS approach can be expressed as a mapping from n-point functions $\Gamma^{(n)}$ to new objects $\bar{\Gamma}^{(2)}_\theta$ and $\bar{\Gamma}^{(2)}_{\theta \theta}$. It bypasses the intermediate infinities encountered in traditional computation. All Feynman integrations give finite results: no divergence (or order $M^2$ contribution) appears. Using the CS equations, \emph{the same conditions as in the textbook case} and finite tree level results, one arrives at the same results for the renormalised two-point functions $\bar{\Gamma}^{(2)}$.} \label{CSfig}
\end{figure}

With that, we understand that the CS method is recursive. Order by order, one can recover the usual results (up to all orders) for  $n$-point functions, but in a manifestly finite way. Contrary to the discussion in Section \ref{standard}, there are neither infinities nor fine-tunings, see Fig. \ref{CSfig}.  All connected $n$-point diagrams can be addressed. Moreover, if the process at the typical energy $E$ is considered, the main contribution to it is coming from the virtual integration momenta $\sim E$. There is no dependence on formally infinite loop momenta, hence there are no intermediate divergences to subtract in the process of computation. In the absence of divergences, no regularisation or renormalisation is required.

\subsection{Absence of fine-tunings for theory with well-separated mass scales\label{CS2}}

Having seen in the previous subsection that the QFT can be formulated in such a way that no infinities appear at any stage of the computation, we can now look at what happens in the theory with two well-separated mass scales. 

We return to the Lagrangian of eq.~(\ref{L2}):
\be
\L = - \frac{1}{2} \left(\d_\mu \phi\right)  \left(\d^\mu \phi\right)- \frac{1}{2} \left(\d_\mu \Phi\right)  \left(\d_\mu \Phi\right) - \frac{m^2}{2} \phi^2- \frac{M^2}{2} \Phi^2 - \frac{\lambda_\phi}{4!} \phi^4 - \frac{\lambda_{\phi\Phi}}{4} \phi^2\Phi^2 - \frac{\lambda_{\Phi}}{4!} \Phi^4~. \label{l2}
\ee
Again, all parameters in it -- $m^2,~M^2,~\lambda_\phi,~\lambda_{\phi\Phi},$ and $\lambda_{\Phi}$ -- are {\em finite} quantities. To make meaningful perturbation expansions, we have to assume, again, that the three four-point couplings $\lambda_\phi$, $\lambda_{\phi\Phi}$ and $\lambda_\Phi$ are all of the same order, which we will denote by just $\lambda$.

Focussing on the two-point functions of the light field $\phi$ and the heavy field $\Phi$, it is clear that at tree level we have
\ba
\left[\bar{\Gamma}^{(2\phi)}\right]_{\lambda^0} &=& i\left(k^2+m^2\right)~,\nn\\
\left[\bar{\Gamma}^{(2\Phi)}\right]_{\lambda^0} &=& i\left(k^2+M^2\right)~.
\ea
Now we turn to the CS approach to the computation of the quantum corrections to the correlation functions in this model. In particular, we focus on the two-point correlation function of the light field, $\bar{\Gamma}^{(2\phi)}$. Our main question is whether there are large corrections $\propto M^2$ to it like those in eq. (\ref{msb}).

In this two-field model, there are two kinds of $\theta$-operations. $\theta_m$ cuts a $\phi$-propagator in two, and $\theta_M$ cuts a $\Phi$-propagator in two. See figure \ref{2th}. In both cases, the newly induced ``cross-vertex" comes with the Feynman rule $(-1)$. From a manifestly finite computation, we find at one loop
\ba
\left[\bar{\Gamma}^{(2\phi)}_{\theta \theta,mm}\right]_\lambda &=& -\frac{i\lambda_\phi}{32\pi^2}~\frac{1}{m^2}~,\nn\\
\left[\bar{\Gamma}^{(2\phi)}_{\theta \theta,mM}\right]_\lambda &=&0~,\nn\\
\left[\bar{\Gamma}^{(2\phi)}_{\theta \theta,MM}\right]_\lambda &=& -\frac{i\lambda_{\phi\Phi}}{32\pi^2}~\frac{1}{M^2}~. \label{gtt2}
\ea

 \begin{figure}[!h]
\begin{center}
\begin{tikzpicture}
[line width=1.5 pt, scale=1.5]

\begin{scope}
\node at (0,0) {$\bar{\Gamma}^{(2\phi)}$};
\end{scope}

\begin{scope}
\node at (0.6,0) {$=$};
\end{scope}

\begin{scope}
[shift={(1,0)}]
\draw (0,0)--(0.75,0);
\node at (1.125,0) {$-$};
\end{scope}

\begin{scope}
[shift={(3,0)}]
\draw (0,0)--(1,0);
\draw [fill=black](0.5,0) circle (0.05cm);
\draw (0.5,0.25) circle (0.25cm);
\node at (1.25,0) {$-$};
\end{scope}

\begin{scope}
[shift={(5,0)}]
\draw (0,0)--(1,0);
\draw [fill=black](0.5,0) circle (0.05cm);
\draw [dashed] (0.5,0.25) circle (0.25cm);
\node at (2,0) {$+~~\O\left(\lambda^2\right)$};
\end{scope}

\begin{scope}
[shift={(0,-1)}]
\node at (0,0) {$\bar{\Gamma}^{(2\phi)}_{\theta,m}$};
\end{scope}

\begin{scope}
[shift={(0,-1)}]
\node at (0.6,0) {$=$};
\end{scope}

\begin{scope}
[shift={(1,-1)}]
\node at (0.375,0) {$1$};
\node at (1.125,0) {$-$};
\end{scope}

\begin{scope}
[shift={(3,-1)}]
\draw (0,0)--(1,0);
\draw [fill=black](0.5,0) circle (0.05cm);
\draw (0.4,0.4)--(0.6,0.6);
\draw (0.4,0.6)--(0.6,0.4);
\draw (0.5,0.25) circle (0.25cm);
\node at (4,0) {$+~~\O\left(\lambda^2\right)$};
\end{scope}

\begin{scope}
[shift={(0,-2)}]
\node at (0,0) {$-\bar{\Gamma}^{(2\phi)}_{\theta,M}$};
\end{scope}

\begin{scope}
[shift={(0,-2)}]
\node at (0.6,0) {$=$};
\end{scope}

\begin{scope}
[shift={(5,-2)}]
\draw (0,0)--(1,0);
\draw [fill=black](0.5,0) circle (0.05cm);
\draw(0.4,0.4)--(0.6,0.6);
\draw(0.4,0.6)--(0.6,0.4);

\draw [dashed] (0.5,0.25) circle (0.25cm);
\node at (2,0) {$+~~\O\left(\lambda^2\right)$};
\end{scope}

\begin{scope}
[shift={(0,-3)}]
\node at (-0.05,0) {$-\bar{\Gamma}^{(2\phi)}_{\theta\theta,mm}$};
\end{scope}

\begin{scope}
[shift={(0,-3)}]
\node at (0.6,0) {$=$};
\end{scope}

\begin{scope}
[shift={(3,-3)}]
\draw (0,0)--(1,0);
\node at (-0.25,0) {$2~\times$};
\draw [fill=black](0.5,0) circle (0.05cm);
\draw (0.4,0.4)--(0.6,0.6);
\draw (0.4,0.6)--(0.6,0.4);
\draw (0.15,0.15)--(0.35,0.35);
\draw (0.35,0.15)--(0.15,0.35);
\draw (0.5,0.25) circle (0.25cm);
\node at (4,0) {$+~~\O\left(\lambda^2\right)$};
\end{scope}

\begin{scope}
[shift={(0,-4)}]
\node at (-0.05,0) {$-\bar{\Gamma}^{(2\phi)}_{\theta\theta,MM}$};
\end{scope}

\begin{scope}
[shift={(0,-4)}]
\node at (0.6,0) {$=$};
\end{scope}

\begin{scope}
[shift={(5,-4)}]
\draw (0,0)--(1,0);
\node at (-0.25,0) {$2~\times$};
\draw [fill=black](0.5,0) circle (0.05cm);
\draw (0.4,0.4)--(0.6,0.6);
\draw (0.4,0.6)--(0.6,0.4);
\draw (0.15,0.15)--(0.35,0.35);
\draw (0.35,0.15)--(0.15,0.35);
\draw[dashed] (0.5,0.25) circle (0.25cm);
\node at (2,0) {$+~~\O\left(\lambda^2\right)$};
\end{scope}

\begin{scope}
[shift={(0,-5)}]
\node at (-0.05,0) {$-\bar{\Gamma}^{(2\phi)}_{\theta\theta,mM}$};
\node at (0.6,0) {$=$};
\node at (7,0) {$~~\O\left(\lambda^2\right)$};
\end{scope}

\end{tikzpicture}

\caption{$(\theta,m)$ and $(\theta,M)$ operations on the one loop correction to the two-point function.\label{2th}}

\end{center}
\end{figure}

In our companion paper \cite{SanderMisha1}, we show that the Callan-Symanzik method can easily be generalised to the case of $n$ interacting scalar fields. We leave the explicit form of the two $n\times n$ matrix equations that generalise eqs.~(\ref{calrel}) to that paper. Here we will just outline the procedure to get extract the desired two-point function from these two equations. 

First, the boundary conditions in eqs.~(\ref{bounren})  get generalised to
\ba
\left[\frac{d}{dk^2}~\bar{\Gamma}^{(2\phi)}(k^2)\right]_{k^2 = 0} =&i&~, \qquad  \qquad \bar{\Gamma}^{(2\phi)}\left(k^2 = 0\right) =i m^2~,\nn\\
\left[\frac{d}{dk^2}~\bar{\Gamma}^{(2\Phi)}(k^2)\right]_{k^2 = 0} =&i~,& \qquad  \qquad \bar{\Gamma}^{(2\Phi)}\left(k^2 = 0\right) =i M^2~.\label{boun2}
\ea
Again, we have chosen to formulate the boundary conditions at $k^2=0$, and again, the method would in principle work as well for any other choice. For our purposes, however, we want this boundary scale to be smaller or equal to (the order of) the mass parameter $m^2$.

Next, evaluating the two-by-two matrix equations for two $\phi$ fields and for two $\Phi$ fields yields the  boundary conditions for $\bar{\Gamma}^{(2\phi)}_{\theta,m}$ and $\bar{\Gamma}^{(2\phi)}_{\theta,M}$ (compare with eq.~(\ref{gtt2})):
\ba
\bar{\Gamma}^{(2\phi)}_{\theta,m}(k^2=0) &=& 1~,\nn\\
\bar{\Gamma}^{(2\phi)}_{\theta,M}(k^2=0) &=& 0. \label{boun3}
\ea
Just like in eqs.~(\ref{gtt3a}), these two results suggest the final ingredients that define the CS approach to this two-field model:
\ba
\left[\bar{\Gamma}^{(2\phi)}_{\theta,m}\right]_{\lambda^0} &=& 1~,\nn\\
\left[\bar{\Gamma}^{(2\phi)}_{\theta,M}\right]_{\lambda^0} &=& 0.\label{bijnakl}
\ea

With that, we have all the necessary input to the two-field generalisation of the system of equations in eqs.~(\ref{calrel}). Their derivation and final form are in our companion paper \cite{SanderMisha1}. Here we just mention that they are two-by-two matrix equations, mixing both $\theta$-operations. As in the one-field case, they allow for a step-by-step, recursive derivation of all two- and four-point functions. As byproducts, they produce the two-field equivalents of the quantities $\beta$, $\gamma$ and $\gamma_\theta$.

We note that, just like in the one field case, there is no dependence on the external momentum in eqs.~(\ref{gtt2}). That is enough to understand that again, there will be no order $\lambda$ contribution to the two-point function of the light field $\bar{\Gamma}^{(2\phi)}$. 

The first corrections to $\bar{\Gamma}^{(2\phi)}$, therefore, occur at order $\lambda^2$. Apart from the contributions depicted in figure \ref{newdeg}, there are new contributions stemming from the sunset diagram with two heavy $\Phi$ propagators. See figure \ref{heavysunset}.

\begin{figure}[!h]
\begin{center}
\begin{tikzpicture}
[line width=1.5 pt, scale=1.5]

\begin{scope}
\draw (0,0)--(1,0);
\draw [fill=black](0.25,0) circle (0.05cm);
\draw [fill=black](0.75,0) circle (0.05cm);
\draw[dashed] (0.5,0) circle (0.25cm);
\draw (0.4,0.15)--(0.6,0.35);
\draw (0.4,0.35)--(0.6,0.15);
\draw (0.4,-0.35)--(0.6,-0.15);
\draw (0.4,-0.15)--(0.6,-0.35);
\node at (-1.5,0) {$ -\bar{\Gamma}^{(2\phi)}_{\theta\theta}~~\supset~~$};
\node at (-0.5,0) {$ 2~~\times$};
\node at (1.5,0) {$ +$};
\end{scope}

\begin{scope}
[shift={(2.75,0)}]
\draw (0,0)--(1,0);
\draw [fill=black](0.25,0) circle (0.05cm);
\draw [fill=black](0.75,0) circle (0.05cm);
\draw (0.5-0.1+0.15,0.19-0.1)--(0.5+0.1+0.15,0.19+0.1);
\draw (0.5+0.1+0.15,0.19-0.1)--(0.5-0.1+0.15,0.19+0.1);
\draw (0.5+0.1-0.15,0.19-0.1)--(0.5-0.1-0.15,0.19+0.1);
\draw (0.5-0.1-0.15,0.19-0.1)--(0.5+0.1-0.15,0.19+0.1);
\draw[dashed] (0.5,0) circle (0.25cm);
\node at (-0.5,0) {$ 4~~\times$};
\node at (1.5,0) {$ +$};
\end{scope}
\end{tikzpicture}

\vspace{1cm}

\begin{tikzpicture}
[line width=1.5 pt, scale=1.5]
\begin{scope}
[shift={(5.5,0)}]
\draw (0,0)--(1,0);
\draw [fill=black](0.25,0) circle (0.05cm);
\draw [fill=black](0.75,0) circle (0.05cm);
\draw[dashed] (0.5,0) circle (0.25cm);
\draw (0.4,0.15)--(0.6,0.35);
\draw (0.4,0.35)--(0.6,0.15);
\draw (0.4,-0.1)--(0.6,0.1);
\draw (0.4,0.1)--(0.6,-0.1);
\node at (-0.5,0) {$ 2~~\times$};
\node at (1.5,0) {$ +$};
\end{scope}

\begin{scope}
[shift={(8.25,0)}]
\draw (0,0)--(1,0);
\draw [fill=black](0.25,0) circle (0.05cm);
\draw [fill=black](0.75,0) circle (0.05cm);
\draw[dashed] (0.5,0) circle (0.25cm);
\draw (0.3,-0.1)--(0.5,0.1);
\draw (0.3,0.1)--(0.5,-0.1);
\draw (0.5,-0.1)--(0.7,0.1);
\draw (0.5,0.1)--(0.7,-0.1);
\node at (-0.5,0) {$ 2~~\times$};
\end{scope}

\end{tikzpicture}
\caption{Two loop contributions to $\bar{\Gamma}^{(2\phi)}_{\theta\theta}$ .\label{heavysunset}}
\end{center}
\end{figure}

The contribution from the diagrams in figure \ref{heavysunset} is most interesting to us because it is here where the large mass $M$ of the field $\Phi$ shows up. From the same system of equations and the same boundary conditions, one finds that the largest contribution to the order $\lambda^2$ part of $\phi$'s correlation function goes as
\be
\left[\bar{\Gamma}^{(2\phi)}\right]_{\lambda^2} \supset \lambda^2\times M^2 \times c_1 \times \ln{\left[  1+c_2\frac{k^2}{M^2}  \right]}.
\ee
Here $c_1$ and $c_2$ are complicated (but convergent) functions of (integrations over) Feynman parameters. They are independent of both masses and external momentum.

Since in this toy model the scale $M$ is associated with the energy scale of new heavy physics, it is safe to assume that experiments will always be restricted to external momenta $k^2 \ll M^2$. Expanding the logarithm we, therefore, end up with
\be
\left[\bar{\Gamma}^{(2\phi)}\right]_{\lambda^2} \supset \lambda^2\times c_1 \times c_2\times \left[k^2+\O\left(k^4/M^{2}\right)\right],\label{expa}
\ee
to be compared with, for instance, the MSbar scheme, where the large corrections appear at first loop order already, see eq.~(\ref{msb}). 

Can we be sure that the absence of large corrections holds up to any loop order? We need two observations to get a positive reply. First, the boundary conditions formulated in eqs.~(\ref{boun2}) and (\ref{boun3}) hold at every loop order. Second, all correlation functions are analytic in $k^2$. That is enough to understand that the absence of large loop corrections persists in all orders of perturbation theory.

Clearly,  in the ``physically accessible" region $k^2/M^2 \ll 1$, all reference to the large scale $M$ scales at most as order $M^0$. In this region, there simply is no order $M^2$ contribution to the two-point function of the light field.  No tunings (cancellations between two large numbers, as is common for multiplicative renormalisation) are needed to keep the mass $m$ much smaller than $M$.

We note that a discussion in terms of the full quantum effective potential $V$ will lead to precisely the same conclusions. Now the boundary conditions, corresponding to  (\ref{boun2}) are imposed on $V$ and have the form
\ba
\frac{\d^2 V}{\d \phi^2}\Big|_{\phi=\Phi=0} &=& m^2, \qquad \frac{\d^2 V}{\d \Phi^2}\Big|_{\phi=\Phi=0} = M^2,\nn\\
\frac{\d^4 V}{\d \phi^4}\Big|_{\phi=\Phi=0} &=&\lambda_\phi,\qquad
\frac{\d^4 V}{\d \Phi^4}\Big|_{\phi=\Phi=0} =\lambda_\Phi,\qquad
\frac{\d^4 V}{\d \phi^2\d\Phi^2}\Big|_{\phi=\Phi=0} =\lambda_{\phi \Phi}~.
\ea
They are symmetric under the substitution $\phi\to\Phi$, and $m\to M$. The explicit finite construction of the effective potential with these boundary conditions is sketched in Subsection \ref{effp} (for the one field case) and in the next publication, \cite{SM2} (for the two field case, including spontaneous symmetry breaking). For now, it suffices to say that the boundary conditions above ensure that the influence of the heavy sector on the light field is negligible. And, again, the CS method does not need any subtraction or renormalisation to arrive at the result that the light field remains light.

From the existence of the finite QFT that evades the corrections of the order $M^2$, we need to conclude that these contributions are simply the artefacts of commonly used multiplicative renormalisation. Applying our toy model to the real world, we deduce that keeping the observed ``small'' Higgs mass to be small does not require dramatic cancellations between two quantities proportional to the scale of new physics $M$ in the divergence-less QFT. 

To summarise, we looked at the two-point function $\bar{\Gamma}^{(2,0)}$ of a light scalar field, coupled to a heavy scalar field. Two different paths (the textbook ``UV divergent" path and the CS ``finite" path) lead to the same result for $\bar{\Gamma}^{(2,0)}$. Of course, this equality only holds if one uses the same renormalisation/boundary conditions for both paths. If we change these conditions that apply to both paths, then we get a new result for $\bar{\Gamma}^{(2,0)}$. (But again, this result comes out regardless of the chosen path.)  As for the influence of the heavy sector on the light one, the boundary conditions (\ref{boun2}) for the ``UV divergent" path do require fine-tunings between the tree values of the Lagrangian parameters and counterterms, whereas in the CS ``finite" path no fine-tunings are needed. This shows that the infinities and fine-tunings in the ``UV divergent" path are unphysical.

\subsection{Sensitivity to the boundary conditions in finite approach to QFT\label{sens}}
As we have seen in this Section, the necessity of fine-tuning of tree parameters of the theory to radiative corrections is formalism-dependent. Even in the framework of generic schemes of multiple renormalisation, different types of subtraction procedures lead to different patterns. For instance, in the single field case of Section \ref{1fmult}, the cutoff regularisation does require that the bare scalar mass $m^2$ diverges as $\Lambda^2$, whereas in dimensional regularisation and MSbar scheme the counter-term vanishes when $m^2\to 0$. 

The conclusion that the fine-tunings are artefacts of the chosen formalism and the scheme of computations can also be seen entirely within a wider generalisation of the CS method. In this work, we use the term ``CS method" to refer to theories in which the boundary conditions are taken at $k^2=0$. For the one field case, these boundary conditions are in eqs.~(\ref{bounren}) and (\ref{gtt2a}). For the two field case, the (relevant) boundary conditions are in eqs.~(\ref{boun2}) and (\ref{bijnakl}). (We remind the reader that all boundary conditions are valid up to all orders in perturbation theory.) 

Let us now consider a ``generalised-CS method", in which the boundary conditions are imposed at some non-zero value $k^2=\Lambda^2$:
\be
\left[\frac{d}{dk^2}~\bar{\Gamma}^{(2)}(k^2)\right]_{k^2 = \Lambda^2} =i,~~   \bar{\Gamma}^{(2)}\left(k^2 = \Lambda^2\right) =i (\Lambda^2+m^2),~~  \bar{\Gamma}^{(4)}\left(k^2 = \Lambda^2\right) =-i\lambda~.
\label{bounrenlambda}
\ee
 To find the pole mass one should iterate the CS equations (\ref{calrel}). If $ (\lambda/16\pi^2)^2 \Lambda^2 \lesssim m^2$ (we remind that the one-loop correction to the scalar propagator is momentum independent, which explains the second power of $\lambda$) the pole mass remains to be close to the tree value $m^2$. If, on the contrary, $ (\lambda/16\pi^2)^2 \Lambda^2 \gg m^2$, the pole scalar mass will be shifted by the two-loop radiative correction by an amount $\delta m^2 \sim (\lambda/16\pi^2)^2 \Lambda^2$. 
 
 If we want to find the same correlation functions as those defined by (\ref{bounren}), the boundary conditions (\ref{bounrenlambda}) must be modified in an obvious way. Take the theory that obeys (\ref{bounren}) and compute in all orders of perturbation theory the vertex functions  $\bar{\Gamma}^{(2)}$ and $\bar{\Gamma}^{(4)}$, evaluate them at momenta $k^2=\Lambda^2$, and use the results to replace the right-hand sides of (\ref{bounrenlambda}). 
 
 This will produce a ``fine-tuned'' expression for the second boundary condition in (\ref{bounrenlambda}): instead of $m^2+\Lambda^2$ one must use $m^2 + \Lambda^2 +{\cal O}(\lambda^2\Lambda^2)+\mbox{higher~orders}$, where the role of corrections is to explain why the light field is light, i.e., to remove the large contribution to the pole mass $\propto \Lambda^2$. Incidentally, a similar consideration applied to the simplest SUSY theory of one scalar superfield (see, e.g. \cite{Wess:1992cp}), considered to be ``natural'', reveals the same necessity of the fine-tuning. Note, however, that here the fine-tuning is between tree-level contributions and loop contributions. The crucial difference between the traditional renormalisation method and the CS method (plus all its generalisations) is that the latter does not run into any fine-tunings between bare correlation functions and counterterms, simply because there is no notion of ``bare", nor of counterterms.

The story above repeats itself in the theory of two scalar fields considered in Section \ref{CS2}. We can change the momentum normalisation scale from $k^2=0$ in (\ref{boun2}) to  $k^2= \Lambda^2$. For $ (\lambda_\phi/16\pi^2)^2 \Lambda^2 \lesssim m^2$ the light field pole mass automatically remains to be close to $m^2$. In the opposite case, the second condition in (\ref{boun2}) must be fine-tuned to get the light, $m^2 \ll \Lambda^2$, scalar field. 

As we will explain below, this observed sensitivity of fine-tunings to the details of the formalism (the boundary conditions in this case) is yet another manifestation of our reasoning that the ``technical'' hierarchy problem is not a physical one.

How to reconcile the conflicting conclusions coming from different formalisms? The answer is simply that in all approaches, $m$, $M$ and $\lambda$ are just Lagrangian parameters, devoid of physical meaning. Only in a  tree-level analysis, they are directly related to physical observables. As soon as loop corrections are included, they become mere tools. Their job is to convert a finite number of initial measurements into predictions for new measurements of physical observables, like particle lifetimes and cross sections\footnote{See for example the lecture notes \cite{Manohar:2018aog}.}. Now, since the Lagrangian parameters do not carry any physical meaning, neither does an alleged fine-tuning between them. The ``unphysicalness" of such a fine-tuning is precisely proven and illustrated by the existence of the CS method: it does not require any fine-tuning but still arrives at the same predictions for physical experiments.

Some time ago one of the authors of this paper (MS) expressed an opinion \cite{Shaposhnikov:2007nj} that the hierarchy and fine-tuning problems are absent in the theories containing no very heavy particles. The intuition there was based on the multiplicative renormalisation in the MSbar scheme. Similar ideas were developed in \cite{Vissani:1997ys,Farina:2013mla}. Our current work states that this argument is unphysical: even if these heavy particles exist, no fine-tunings are needed in finite QFT.

\section{``Naturalness'' and ``calculability'' of the Higgs mass.\label{calc}}

As announced in the Introduction, our paper focuses on the ``technical" hierarchy problem. We show that there is no need to worry about fine-tunings between tree parameters and loop corrections because these fine-tunings are formalism dependent and thus unphysical. To back up that statement, we have presented another computation of renormalised correlation functions in which such fine-tuning does not occur. Actually, there is no notion of renormalisation at all in the CS method.

\red{
The two-field model that we have used so far to illustrate the use of the CS method is an example of a theory in which the Appelquist - Carrazzone decoupling theorem clearly holds. At the end of the day, the low-energy sector of the theory (in this case, the physics of the light field $\phi$) is not sensitive to its high-energy sector (in this case, the physics of the field $\Phi$). This conclusion holds irrespective of whether one uses traditional renormalisation or the CS method. In this way, one could say that the CS method provides an explicit illustration of the decoupling theorem.
}

However, there exists another, related version of the problem -- in this paper referred to as the ``hierarchy problem". Take a UV complete theory containing a mass scale $\Lambda$ much larger than the Fermi scale and which has fewer free parameters than the SM. In this theory, the Higgs mass can potentially be predicted. It is \red{widely} argued \red{(see, for example, the lecture notes \cite{Cohen:2019wxr} and references therein)} that in this setup the Higgs mass is ``fed'' generically by this high energy scale, bringing it to $\Lambda$ unless there is precise fine-tuning between the parameters of the UV theory. \red{Note that now, the fine-tuning occurs in the already renormalised theory. This is to be contrasted with the fine-tuning that we have discussed so far (i.e., the technical hierarchy problem), which occurs \emph{during} the traditional approach to renormalisation, namely in the balancing of \emph{bare} loop corrections and counterterms.} 

In this Section, we explain why to us, this \red{version of the} problem is \red{still} as unphysical as the ``technical" hierarchy problem addressed in this paper. \red{Admittedly, while in the previous sections,  we could rely on firm calculations, this section is somewhat more ``personal". It contains our point of view on the hierarchy problem. For sure, other authors could argue in the opposite way and not share this point of view (which by itself is a motivation to include this section). For example,  the lecture notes \cite{Cohen:2019wxr}  provide a wording of the dominant, quite a different take on the problem.}

\subsection{What is the UV theory?}

Needless to say, the ultimate theory of quantum gravity, unifying general relativity and quantum field theory, has not been constructed yet. At a slightly less ambitious level, the Standard Model, perhaps, could merge in a UV complete theory below the Planck scale. Can we be sure that a UV completion of the Standard Model described by less than $19$ parameters must exist?  We do not have yet any hint in favour of this hypothesis. If there is no such a theory, there is no notion of computing the Higgs mass from it, and therefore no hierarchy problem. We will comment further on this in the final section of this paper.

\subsection{UV choice of Lagrangian parameters}

The alleged necessity of fine-tuning in the UV theory again occurs between unphysical Lagrangian parameters.  Just like in the ``technical" version of the problem, Lagrangian parameters are unphysical tools. These are nothing but instruments to tie results and predictions for real, physical experiments together. Different choices of formalism to do computations of radiative corrections give different values for Lagrangian parameters but do produce the same physical predictions. Lagrangian parameters are enormously valuable tools but do not deserve any worries about their actual values. Moreover, the unique set of UV parameters (if there is more than one of them) cannot be selected unambiguously. For instance,  in the case of a SU(5) GUT, computing the Higgs mass in terms of the masses of the scalar $\underline{5}$  and $\underline{24}$  multiplets, needs fine-tuning, whereas computing it in terms of the vacuum expectation values of these fields does not.

\subsection{``Feeding'' the Higgs mass}

According to \cite{Cohen:2019wxr}, the hypothetical UV theory has a hierarchy problem ``unless there is a protection mechanism, preventing the new high scales feed into the Higgs mass parameter''. The word ``feed'' is quite vague and does not seem to allow an exact definition. It is the model and formalism dependent. For example, in the two-field model considered in detail in our work, the heavy scalar mass ``feeds''  into the mass of the light scalar field if the multiplicative renormalisation in the MSbar scheme is used, but does not ``feed'' it in the CS approach. Note that only the imaginary parts of the Green's functions are formalism independent, whereas the terms under discussion are real and have nothing to do with absorptive parts of the Green's functions. In other words, the ``hierarchy problem'' is not different from the ``technical hierarchy problem'' in this respect.

\subsection{Naturalness predictions}

The belief in the power of naturalness considerations receives enormous boosts from the success of its alleged predictions in the context of effective field theories.  Here we first sketch the standard argument, and then we comment on it.

In standard reasoning, one computes quantum corrections to the measured (small) mass of some scalar particle within the cutoff regularisation. This gives a result proportional to $\Lambda^2$, where $\Lambda$ is the cut-off of the loop integral. This cut-off is now given a physical interpretation: it is the scale where an underlying UV theory begins to manifest itself. Naturalness then prescribes that such loop corrections can at most be as large as the measured number itself, but not larger. That would be unnatural: to explain a measured number, two much larger numbers (tree value and loop correction) should almost cancel each other. "Naturally" setting the quantum corrections equal to the measured number then yields the scale of new physics $\Lambda$.

A well-known example involves the mass difference between the charged and neutral pions. This difference receives a quadratically divergent electromagnetic contribution (from a photon loop)  of the order
\be
m_{\pi^+}^2 - m_{\pi^0}^2 = \frac{3 \alpha}{4\pi}\cdot \Lambda^2,
\label{pion}
\ee
with $\alpha$ the fine structure constant and $\Lambda$ is taken as the cut-off at which the theory breaks down. Inserting the small pion mass difference at the left-hand side gives $\Lambda \leq 850$ MeV. And, indeed, before hitting this cut-off, at $770$ MeV one encounters the $\rho$-meson. 

Another example is in the context of mixing between kaon mesons, first pointed out in \cite{Gaillard}. For the mixing between the ``long" and ``short" states of neutral kaons, we also have a quadratically divergent contribution
\be
\frac{M_{K^0_L}-M_{K^0_S}}{M_{K^0_L}} = \frac{G_F^2 f_K^2}{6\pi^2}\sin{\theta_c}^2 \cdot \Lambda^2,\label{kaon}
\ee
with $f_K$ the kaon decay constant and $\theta_c$ the Cabibbo angle. The observed $\mathcal{O}\left(10^{-15}\right)$ ``unnatural" value at the left hand side corresponds to $\Lambda \simeq 2$ GeV. Indeed, at around $1.2$ GeV the charm quark shows up, marking the cut-off of the effective kaon theory and thereby providing an explanation for the small value measured at the left-hand side of the equation.

Playing this same game for the Higgs mass, taking into account loop corrections from the top quark, the $W$ and $Z$ bosons and the Higgs itself gives
\be
\delta m_h^2 = \frac{3 G_F}{4\sqrt{2}\pi^2}\left( 4m_t^2-2m_W^2-m_Z^2 - m_H^2 \right).
\ee
Requiring that this loop correction does not exceed the measured value $m_H=125~{\rm GeV}$ yields the prediction $\Lambda\simeq 500~{\rm GeV}$ as the scale where new physics should come in.

Our comments on this strategy are as follows. To begin with, nature does not do perturbation theory and therefore does not know about ``loop corrections''. The separation between tree-level contributions and loop contributions is purely instrumental. More to the point, the existence of the CS method shows that quadratic divergences are unphysical anyway. Physics is in the mapping of one measurable quantity to another. Using an unphysical, formal and instrumental separation for physics predictions simply comes down to an unjustified use of artificial tools.

Nevertheless, let us assume that in some way or another one reaches an agreement on which method to use to compute loop corrections. Then we are conjectured to be in the position to make the (pre/post)-dictions given above for the mass of the $\rho$ meson, the charm quark and, generalising the logic, the scale of new (supersymmetric?) physics. However, we have seen that this prediction in the end comes down to comparing loop corrections in cutoff regularisation to measured results. It is presumed that ``naturally"  they should be of about the same size. But what is ``about"? Precisely equal? Or do we still accept a factor of five between them? To say that on naturalness grounds, the pion mass difference indicates that $\rho$ meson mass needs to be below 850 MeV and that the charm quark needs to be lighter than 1.2 GeV, one addresses an unjustified precision to the requirement ``natural". (Admittedly, taking $10^{-120}$ as ``unnatural" can probably be well defended. That illustrates the well-known, majestic failure in predicting the value of the cosmological constant through naturalness arguments.)

\subsection{A ``finite view" on naturalness predictions}

Returning to our argument, it is instructive to see how these ``naturalness-based predictions" can be recovered in the CS approach discussed in Section \ref{sens}, and how they are to be understood in this context. Let us take for definiteness the pion example. The low energy theory of pions by itself is renormalisable, and as such, it is valid up to exponentially high energies, related to Landau poles in the coupling constants. The masses of particles in it are the experimental input, and cannot be computed\footnote{Of course, the pion masses can be determined if one uses the underlying theory - QCD.  See also the classic paper \cite{Das} for the computation of the pion mass difference in effective field theory, free from divergences.}. 

Let us now put the boundary conditions for mass operators ${\bar{\Gamma}}^{(2)}_{\pi^{0,+}}(k^2)$ corresponding to the charged and neutral pion at scale $\Lambda$ and require that they are equal to each other, ${\bar{\Gamma}}^{(2)}_{\pi^{0}}(k^2 = \Lambda^2)={\bar{\Gamma}}^{(2)}_{\pi^{+}}(k^2 = \Lambda^2)$ arguing that at this scale the isospin symmetry should be exact. This yields again relation (\ref{pion}), and the same numerical value for $\Lambda$ as previously. In other words, the scale $\Lambda$ can be computed from the observed pion mass differences. However, its physical sense has nothing to do with the scale of new physics - it is simply the normalisation point at which the specific boundary conditions are imposed. The success of the naturalness paradigm with $\pi$ and $K$ mesons is thus not generic and is rooted deeply into the structure of the underlying renormalisable field theory - QCD in this case. It is instructive to compare the QCD and the SM at this point. In QCD (let's leave away electromagnetic and weak interactions for simplicity), the meson masses can be expressed through 7 parameters - $\Lambda_{QCD}$ and 6 quark masses (we leave away the QCD $\theta$ angle). Since the number of all mesons is greater than the number of parameters, the theory is predictive: by making 7 different measurements (for example, the masses of $\rho,\pi^0,\pi^+, K^+, D^0, B^+$, and one containing the top quark) one can predict the results of all experiments with strongly interacting particles. But this is nothing more but the statement that in renormalisable theories characterised by $N$ ($N=7$ for QCD) parameters $N$ measurements allow predicting an infinite number of observables. In the Standard Model, the number of parameters is 18 (this number, of course, is larger than 7), but the conclusion is the same. Therefore, QCD is no better than the Standard Model from the point of view of predictivity.

 Along the same lines, the relation between the Higgs mass in the SM and the parameter $\Lambda$, $m_H^2 \sim f_t^2 \Lambda^2$ in the generalisation of the CS method is associated with the specific boundary condition imposed on the Higgs mass operator at $k^2 = \Lambda^2$. Nothing drastic happens at this energy scale. The Standard Model by itself is perfectly consistent for energies much exceeding the Planck scale without any new physics. From this point of view, it is not surprising that naturalness arguments fail for the electroweak theory (they require the existence of new physics below $1$ TeV) and for the cosmological constant problem (they predict new physics at the scale of a fraction of eV). This failure is telling us that the fundamental theory embracing the Standard Model is most probably not QCD-like.

We end with a small observation. These alleged naturalness predictions do not depend at all on the nature of the ``new physics" that is supposed to appear at energy scale $\Lambda$. Consequently, the requirement that the Higgs mass must be calculable from these new physics does not play a role. It just assumes that such new physics must exist. In this respect,  these predictions apply more to the ``technical hierarchy problem" (i.e., large cancellations between loop corrections and tree values), than to the problem of the fine-tuning occurring when computing the Higgs mass from a more fundamental UV theory.

\section{Cosmological constant and effective potential}
\label{cc}

\subsection{Cosmological constant}

So far, we have focused on avoiding the quadratic and logarithmic UV divergences appearing in the standard computation of two- and four-point functions. Of course, we can compute zero-point functions as well.
Let us consider the one-field model with an extra cosmological constant $\Lambda$ (of mass dimension four) included:
\be
\L = - \frac{1}{2} \left(\d_\mu \phi\right)  \left(\d^\mu \phi\right)-\Lambda- \frac{m^2}{2} \phi^2 - \frac{\lambda}{4!} \phi^4 ~. \label{lageff}
\ee
At the tree level, we simply get 
\be
\Gamma^{(0)}_{\rm tree}= -i\Lambda~.
\ee
Including one loop corrections (i.e., vacuum bubble diagrams, see figure \ref{cosmcon}) cut off  at the formal scale $\Lambda_{\rm cut-off}$ (i.e., the formal scale that we called $\Lambda$ before), the standard computation yields a quartic divergence:
\be
\Gamma^{(0)}= -i\Lambda+c\times \Lambda_{\rm cut-off}^4+\O\left(\lambda\right)~,
\ee
with $c$ some numerical constant.

\begin{figure}[!h]
\begin{center}
\begin{tikzpicture}
[line width=1.5 pt, scale=1.5]

\begin{scope}
[shift={(0.25,0)}]
\draw (0.5,0.25) circle (0.25cm);
\node at (-0.5,0.25) {$\Gamma^{(0)}~~=$};
\node at (1.5,0.25) {$+~~\O\left(\lambda\right)$};
\end{scope}

\end{tikzpicture}
\caption{First (order $\lambda^0$) quantum correction to $\Gamma^{(0)}$.\label{cosmcon}}
\end{center}
\end{figure}

Now, in practice one measures a very small cosmological constant. Therefore, the standard approach requires tremendous fine-tuning between the two contributions above. However, this is again a fine-tuning between unphysical Lagrangian parameters, so there is nothing to worry about. The only physical prediction is that since there is by construction no dependence on any external momenta in these zero-point diagrams, measurements of the cosmological constant will always return the same value.

Here we want to show that the reasoning above can be supported further by an extension of the CS method. For the zero-point function, we need three derivatives with respect to the mass parameter to arrive at finite results. In other words, we need a third equation, to compute the convergent quantity $\bar{\Gamma}^{(0)}_{\theta\theta\theta}$. See figure \ref{cosmcontheta}.   

However, with the new Lagrangian parameter $\Lambda$ in the problem, the first and second equation change as well. We postulate
\ba
2 m^2 ~(1+\gamma)~\cdot i\cdot\bar{\Gamma}^{(n)}_\theta&=&  \left(m ~\frac{\d}{\d m} +\beta~\frac{\d}{\d \lambda}  +\gamma_\Lambda ~\frac{\d}{\d \Lambda}   +n\cdot \gamma\right) \bar{\Gamma}^{(n)}~,\nn\\
2m^2~(1+\gamma) \cdot i\cdot  \bar{\Gamma}^{(n)}_{\theta \theta} &=& \left(m~\frac{\d}{\d m} +\beta~\frac{\d}{\d \lambda}+\gamma_\Lambda ~\frac{\d}{\d \Lambda} +n\cdot\gamma+\gamma_\theta\right)\bar{\Gamma}^{(n)}_\theta~,\nn\\
2m^2~(1+\gamma) \cdot i\cdot  \bar{\Gamma}^{(n)}_{\theta \theta\theta} &=& \left(m~\frac{\d}{\d m} +\beta~\frac{\d}{\d \lambda}+\gamma_\Lambda ~\frac{\d}{\d \Lambda} +n\cdot\gamma+2\cdot \gamma_\theta\right)\bar{\Gamma}^{(n)}_{\theta\theta}~,\label{3cweq}
\ea
with  $\gamma_\Lambda$ a new quantity of mass dimension four. The anomalous dimension $\gamma_\Lambda$ begins at order $\lambda^0$ already: we can make a loop diagram with zero vertices. (That is precisely the first correction to the zero-point function.)

From this system, one can again follow the recursive approach of Section \ref{Cal}. As for the computation for the two-and four-point functions, nothing changes. Neither the corresponding convergent diagrams $\bar{\Gamma}^{(2)}_{\theta\theta}$ and $\bar{\Gamma}^{(4)}_\theta$, nor their boundary conditions involve the ``zero-point vertex" $\Lambda$. The derivative with respect to $\Lambda$ will always vanish. At this point, the value of $\gamma_\Lambda$ is irrelevant.

Next, we turn to the computation of the zero-point function. By construction, no diagram can depend on external momenta. The choice for boundary conditions at some given scale then directly gives the results at any other scale. Therefore, in a way the CS formalism is empty now: the actual result for the corrections to the zero-point function is completely determined by the chosen boundary conditions at $k^2=0$ (or, equivalently, the chosen integration constants).

However, it is still worth asking whether there exist choices for $\gamma_\Lambda$ that yield a self-consistent set of equations. As an Ansatz for the resulting zero-point function, we take
\be
\bar{\Gamma}^{(0)}= -i\Lambda ~.
\label{gan}
\ee

We will restrict ourselves to the simplest case: the order $\lambda^0$ contribution to the zero-point function. The system of the equation then reduces to
\ba
i \cdot\bar{\Gamma}^{(0)}_\theta&=&  \left(\frac{\d}{\d m^2}  +\frac{\gamma_\Lambda}{2m^2} ~\frac{\d}{\d \Lambda}   \right) \bar{\Gamma}^{(0)}~,\nn\\
i\cdot  \bar{\Gamma}^{(0)}_{\theta \theta} &=& \left(\frac{\d}{\d m^2} +\frac{\gamma_\Lambda}{2m^2} ~\frac{\d}{\d \Lambda} \right)\bar{\Gamma}^{(0)}_\theta~,\nn\\
i\cdot  \bar{\Gamma}^{(0)}_{\theta \theta\theta} &=& \left(\frac{\d}{\d m^2} +\frac{\gamma_\Lambda}{2m^2} ~\frac{\d}{\d \Lambda} \right)\bar{\Gamma}^{(0)}_{\theta\theta}~. \label{cs3}
\ea
Putting the Ansatz of eq.~(\ref{gan}) into the first two equations yields, respectively,
\ba
\bar{\Gamma}^{(0)}_\theta(k^2=0)&=& 
-\frac{\gamma_\Lambda}{2m^2}~,\nn\\
\bar{\Gamma}^{(0)}_{\theta \theta}(k^2=0) &=&
 - \frac{i\cdot\gamma_\Lambda}{2 m^4} +\frac{i}{2m^2}\cdot\frac{\d\gamma_\Lambda}{\d m^2}+\frac{i\cdot\gamma_\Lambda}{4m^4}\cdot \frac{\d \gamma_\Lambda}{\d \Lambda}~.\label{fte}
\ea
To have a clear analogy with the CS computations of the two- and four-point functions, we have written the above equations as boundary conditions at $k^2=0$, which determine the integration constants. Since $\bar{\Gamma}^{(0)}$, $\bar{\Gamma}^{(0)}_\theta$ and $\bar{\Gamma}^{(0)}_{\theta\theta}$ are all independent of the external momenta, at all values for $k^2$ they are all equal to these prescribed values. 

However, at  the left-hand side of the third equation in eqs.~(\ref{cs3}), we encounter the manifestly convergent quantity $\bar{\Gamma}^{(0)}_{\theta \theta\theta} $ which can be computed in the standard, unambiguous way
\be
\bar{\Gamma}^{(0)}_{\theta \theta\theta} = 2\times \frac{1}{2}\times (-1)^3 \int \frac{d^4l}{(2\pi)^4}~\left(\frac{-i}{l^2+m^2}\right)^3 =\frac{1}{32\pi^2}\frac{1}{m^2}~. \label{g3t}
\ee
\begin{figure}[!h]
\begin{center}
\begin{tikzpicture}
[line width=1.5 pt, scale=1.5]

\begin{scope}
\node at (-1,0.25) {$\bar{\Gamma}^{(0)}_{\theta\theta\theta}~~=$};
\node at (-0.25,0.25) {$2~~\times $};
\draw (0.5,0.25) circle (0.25cm);
\draw (0.4,0.4)--(0.6,0.6);
\draw (0.6,0.4)--(0.4,0.6);
\draw (0.4,-0.1)--(0.6,0.1);
\draw (0.6,-0.1)--(0.4,0.1);
\draw(0.15,0.15)--(0.35,0.35);
\draw(0.15,0.35)--(0.35,0.15);
\node at (1.5,0.25) {$+ ~~~\O\left(\lambda\right)$};
\end{scope}

\end{tikzpicture}
\caption{First (order $\lambda^0$) quantum correction to $\bar{\Gamma}^{(0)}_{\theta\theta\theta}$.\label{cosmcontheta}}
\end{center}
\end{figure}

Inserting this result back into the last line in eqs.~(\ref{cs3}) and using the second boundary condition in eqs.~(\ref{fte}) yields a long equation involving many derivatives of $\gamma_\Lambda$ with respect to the parameters $m^2$ and $\Lambda$. Demanding $\gamma_\Lambda$ to be analytic in both these parameters yields the unique solution
$
\gamma_\Lambda = \alpha \cdot m^4+\beta\cdot \Lambda~,
$
where the coefficients $\alpha$ and $\beta$ satisfy
\be
\frac{1}{32\pi^2} = \frac{\alpha\beta(-2+\beta)}{8}+\frac{\beta\Lambda}{ m^4}\left(1-\frac{3\beta}{4}+\frac{\beta^2}{8}\right)~. 
\ee
The solution to this equation is
\be
\alpha = \frac{1}{32\pi^2}~, \qquad \qquad \beta=4~.
\ee
Therefore, we find that the system (\ref{cs3})  is indeed  consistent with the boundary condition (\ref{gan}) with
\be
\gamma_\Lambda = 4\Lambda+\hbar\cdot \frac{m^4}{32\pi^2}~.
\label{gammaflat}
\ee
Here, we have reinstalled factors of $\hbar$ to stress that the first term at the right-hand side arises from tree-level physics, while the second one is generated at the one-loop level.

Again, the CS method cannot generate a prediction for the zero-point function, since the final result directly depends on the chosen boundary conditions (or, equivalently, integration constants). However, from the above, we see that there exists a unique solution for $\gamma_\Lambda$ such that a self-consistent computation yields vanishing loop corrections to the zero-point function.

We end this Subsection with a comment. In flat Minkowski space without gravity the vacuum energy $\bar{\Gamma}^{(0)}$ is not observable. However, if the space is compact or contains ``defects''  leading to non-trivial conditions imposed at the fields at the boundaries, the zero-point function starts to depend on the geometry in question. This is the Casimir effect. The solution of equations (\ref{cs3})  with the condition  (\ref{gan}) imposed for {\em flat Minkowski space} and anomalous dimension given by (\ref{gammaflat})  gives the dependence of the vacuum energy on this geometry. And, again, no infinities or fine-tunings are encountered in the course of this computation.

\subsection{Effective potential \label{effp}}

In the theories we considered up to now the mass of the light scalar is an arbitrary parameter that cannot be expressed via other constants of the theory. If the tree-level low-energy theory has an extra symmetry, such as scale or conformal invariance, the tree value of the scalar mass is zero, but the physical mass is non-zero and can be predicted \cite{Coleman:1973jx}. This setup was used for a possible explanation of the small but non-zero value of the Higgs mass in Grand Unified Theories in \cite{Weinberg:1978ym}, and in \cite{Shaposhnikov:2018xkv,Shaposhnikov:2020geh} for a conceivable relation between the Fermi and Planck scales.

A helpful instrument that allows for the resummation of perturbation theory or the study of non-perturbative effects is associated with the effective potential for the scalar field. (The notion of the effective potential is of course very much related to the cosmological constant/zero-point energy.) For the  theory of eq.~(\ref{lageff}) (the generalisation to a multi-field case is straightforward), the Coleman-Weinberg effective potential \cite{Coleman:1973jx} is defined in terms of $n$-point functions as
\be
-iV_{\rm eff}(\phi_0) =
\sum_n \frac{\phi_0^n}{n!}\cdot\bar{\Gamma}^{(n)}\Big|_{k^2=0}~. \label{veff}
\ee
(To be clear: while $V_{\rm eff}$ by itself is  ($i$ times) a sum of zero-point functions at non-zero scalar field background $\phi_0$, it is expressed here in terms of $n$-point functions.) Here $\phi_0$ denotes the vacuum expectation value of the field $\phi$.

\begin{figure}[!h]
\begin{center}
\begin{tikzpicture}
[line width=1.5 pt, scale=1.5]

\begin{scope}
[shift={(-2.5,1.5)}]
\draw (0.5,0) circle (0.05cm);
\node at (1.65,0) {$ +$};
\end{scope}

\begin{scope}
[shift={(0,1.5)}]
\draw [fill=black] (0,0)--(1,0);
\node at (1.65,0) {$ +$};
\end{scope}

\begin{scope}
[shift={(3,1.5)}]
\draw (-0.3,0.3)--(0.3,-0.3);
\draw (-0.3,-0.3)--(0.3,0.3);
\draw [fill=black](0,0) circle (0.05cm);
\end{scope}

\begin{scope}
[shift={(-2.5,-0.15)}]
\draw (0.5,0.25) circle (0.25cm);
\node at (-0.5,1.65) {$ -iV_{\rm eff}~~~=$};
\node at (1.65,0.15) {$ +$};
\node at (1.85,-1.3) {$ +~~\dots$};
\end{scope}

\begin{scope}
\draw [fill=black] (0,0)--(1,0);
\draw [fill=black](0.5,0) circle (0.05cm);
\draw (0.5,0.25) circle (0.25cm);
\node at (1.65,0) {$ +$};
\end{scope}

\begin{scope}
[shift={(2.5,0)}]
\draw (0,0.25)--(0.25,0);
\draw (0,-0.25)--(0.25,0);
\draw [fill=black](0.25,0) circle (0.05cm);
\draw (0.5,0) circle (0.25cm);
\draw [fill=black](0.75,0) circle (0.05cm);
\draw (0.75,0)--(1,0.25);
\draw (0.75,0)--(1,-0.25);
\node at (1.65,0) {$ +$};
\end{scope}

\begin{scope}
[shift={(5,0)}]
\draw (0,0.25)--(0.25,0);
\draw (0,-0.25)--(0.25,0);
\draw [fill=black](0.25,0) circle (0.05cm);
\draw (0.5,0) circle (0.25cm);
\draw [fill=black](0.75,0) circle (0.05cm);
\draw (0.75,0)--(1,0.25);
\draw [fill=black](0.5,0.25) circle (0.05cm);
\draw (0.5,0.25)--(0.75,0.5);
\draw (0.5,0.25)--(0.25,0.5);
\draw (0.75,0)--(1,-0.25);
\end{scope}

\end{tikzpicture}
\end{center}
\caption{Tree level (first line) and one loop (second line) contributions to the effective potential. All external lines carry the classical field $\phi_0$, all internal lines carry the quantum field $\delta \phi$. \label{gefffig} }
\end{figure}

Clearly, by inserting the expressions for $\Gamma^{(n)}$ that we found in Section \ref{Cal} one can obtain the effective potential in a manifestly finite way. At the tree level, we have contributions for $n=0,2,4$. At the one-loop level, the first contribution appears for $n=6$, since we have imposed ourselves that the zero-, two- and four-point functions vanish at zero external momentum. One arrives at the familiar result
\be
V_{\rm eff} =
\Lambda+\frac{m^2}{2}\phi_0^2+\frac{\lambda}{4!}\phi_0^4
-\frac{\hbar}{64\pi^2}\left[
\frac{\lambda \phi_0^2}{8}\left(4m^2+3\lambda\phi_0^2\right)-\left(m^2+\frac{\lambda\phi_0^2}{2}\right)^2\ln{\left(1+\frac{\lambda \phi_0^2}{2m^2}\right)}
\right] ~.\label{geff}
\ee
(Here we have once more reinserted the factor of $\hbar$ to show explicitly that the effective potential is an expansion in powers of $\hbar$, not in powers of the coupling constant.)

To show the consistency of our approach, it is worth mentioning that there is yet another way to obtain effective potential without running into intermediate divergences. We can again make use of eqs.~(\ref{3cweq}), but with $n$ eliminated in favour of $\phi_0~\d/\d \phi_0$:

\ba
2 m^2 ~(1+\gamma)~\cdot i\cdot V_{{\rm eff},\theta}&=&  \left(m ~\frac{\d}{\d m} +\beta~\frac{\d}{\d \lambda}  +\gamma_\Lambda ~\frac{\d}{\d \Lambda}   + \gamma \cdot \phi_0\frac{\d}{\d \phi_0}\right) V_{{\rm eff}}~,\nn\\
2m^2~(1+\gamma) \cdot i\cdot  V_{{\rm eff},\theta\theta} &=& \left(m~\frac{\d}{\d m} +\beta~\frac{\d}{\d \lambda}+\gamma_\Lambda ~\frac{\d}{\d \Lambda} +\gamma \cdot \phi_0\frac{\d}{\d \phi_0}+\gamma_\theta\right)V_{{\rm eff},\theta}~,\nn\\
2m^2~(1+\gamma) \cdot i\cdot  V_{{\rm eff},\theta\theta\theta} &=& \left(m~\frac{\d}{\d m} +\beta~\frac{\d}{\d \lambda}+\gamma_\Lambda ~\frac{\d}{\d \Lambda} +\gamma \cdot \phi_0\frac{\d}{\d \phi_0}+2\cdot \gamma_\theta\right)V_{{\rm eff},\theta\theta}~.\nn\\
\ea
Now, performing three $\theta$ operations on each of the diagrams in fig.~\ref{gefffig} separately, we find
\be
V_{{\rm eff},\theta\theta\theta}=\frac{i}{32\pi^2 m^2}\left[1-\frac{\lambda\phi_0^2}{2m^2}+\frac{\lambda^2 \phi_0^4}{4m^4}+\dots\right] = \frac{i}{32\pi^2}\frac{1}{m^2+\lambda\phi_0^2/2}~.
\ee
Inserting this result in the system of equations above, with an extra one loop contribution to the quantity $\gamma_\Lambda$:
\be
\gamma_\Lambda = 4\Lambda+\hbar\left(\frac{m^4}{32\pi^2}-\frac{\lambda\Lambda}{8\pi^2}\right)~,
\ee
 we can ``integrate our way up" to obtain the same result for $V_{\rm eff}$ as in eq.~(\ref{geff}).

We note that the existence of the method described above implies that in the limit $m^2\to 0$, corresponding to a scale-invariant theory, we can then compute the scalar mass without any intermediate fine-tuning. No subtractions are needed to satisfy the boundary conditions. This illustrates, once again, that in theories with separate mass scales the UV sector does not influence the mass parameter of the light (say, Higgs) scalar field. We postpone the explicit construction of the two-field effective potential to a further publication \cite {SM2}.

\section{Conclusions}
\label{concl}

\subsection{Summary}

The aim of this work has been to stress that there is no physical sense in the ``technical'' naturalness criterion. ``Fine-tuned cancellations" in the computation of renormalised correlation functions\footnote{Even the renormalised correlation functions by themselves are unphysical. They depend on the chosen renormalisation scheme and involve unphysical Lagrangian parameters. These ambiguities disappear only after combining the renormalised correlation functions with results from some initial measurements. Only then predictions for physical observables as decay rates and cross-sections can be made.} are between unphysical quantities -- Lagrangian parameters. 

To explicitly illustrate this fact, we have looked at an old approach to renormalisation, introduced by Callan in \cite{Callan:1975vs}. Divergent or ``large" subtractions never appear in this method, see Figures \ref{figtrad} and \ref{CSfig}.  Originally intended as an all-order proof of the renormalizability of $\lambda \phi^4$ theory, we have used this old method for a new purpose: an analysis of naturalness issues (a topic that was not discussed at the time when the method was introduced). 

We stress that at the end of the day, we are not interested in actually \emph{using} this CS method. It is much more involved than the textbook approach to renormalisation. All we need to make our point is the mere \emph{existence} of the CS method. If it is possible to compute renormalised correlation functions in a formalism without any intermediate fine-tuning, then this fine-tuning has to be formalism-dependent. That gives additional evidence to the claim made in the first paragraph above.

\subsection{Divergences and fine-tunings}

The necessity of fine-tunings in the theories with several very different mass scales does show up in the perturbative multiplicative renormalisation procedure, used most often in practical computations, as well as in the Wilsonian approach. In whatever renormalisation scheme (cutoff, DimReg, Pauli-Villars, etc), the computations are plagued by UV divergences. The same is true for the non-perturbative lattice approach to QFT, where the lattice spacing generates a physical UV cutoff. All these are the source of the ``technical naturalness'' criterion.

As we discussed in this paper, the appearance of UV divergences, however, is for\-ma\-lism dependent: we have reviewed formulations of QFT that do not have  UV divergences whatsoever. We described how the finite QFT can be constructed based on Callan-Symanzik equations and convergent Feynman diagrams, supplemented by specific boundary conditions. Moreover, we have shown that this method does not require any large (but finite) fine-tunings either in the theories with well-separated mass scales. The quantum corrections to physics at the energy scale $E$ are dominated by the momenta of the same order, since all expressions are given by convergent integrals, in full accordance with the Appelquist-Carazzone theorem \cite{Appelquist:1974tg}. 

\subsection{Naturalness}

This shows that the so-called technical hierarchy problem (the sensitivity of low-energy physics to high-energy physics) depends on the formulation of quantum field theory, and, therefore, is devoid of physical meaning, at least for renormalisable theories. On a very related note, it is often said that while in classical theories small and large numbers can perfectly coexist together, this does not hold for quantum theory. For example, in the standard approaches to renormalisation it seems that when large masses are present, small masses cannot exist because unless dramatic fine-tuning is imposed, quantum corrections tend to lift them to the order of magnitude of the large masses. Here we have argued that, on the contrary, there is no difference between classical field theory and quantum field theory in this respect. The problem of stability of the electroweak scale or the cosmological constant against quantum corrections is just an artefact of the commonly chosen computation method, based on perturbative Feynman graphs and renormalisation. Fine-tunings are absent in the divergent-free formulation. In short, this paper does not solve the hierarchy problem, it rather shows that it does not exist.

We are not the first ones who argue with the technical naturalness criterion. A nice description of the hierarchy problem from the point of view of effective field theories can be found in the lectures by A. Manohar \cite{Manohar:2018aog}, and we agree with the point of view advocated there. A close discussion is presented in \cite{Neumaier}. Just like we have argued here, these authors stress that Lagrangian parameters do not carry physical meaning by themselves. One should not be worried by cancellations, small, large or formally infinite, between physically meaningless parameters. 

\red{We note that in the closely related ``hierarchy problem" the setting is a bit different. In that context, one is famously concerned with the fine-tuning occurring when computing the Higgs mass out of a UV completion of the Standard Model. In Section \ref{calc} we have explained why, from our point of view, not quite in line with the mainstream in our community, this problem of fine-tuning does not essentially differ in spirit from the technical hierarchy problem of fine-tuning. }

\subsection{Generalisations}
Our discussion can be extended in several directions. We considered only one possibility of constructing the divergence-free QFT, based on CS equations. It would be interesting to see what happens with ``technical naturalness'' in the BPHZ (or its modifications) and 't Hooft \cite{tHooft:2004bkn} formulations of finite QFT. Also, we have only dealt with renormalisable field theories. It looks obvious that the system of recursive CS equations can be generalised to the non-renormalisable case as well by repeating the ``$\theta$-operation'' as many times as needed.  We expect that the qualitative conclusions about the absence of fine-tunings will remain the same even in this case. On the technical side, finally,  the CS method as it stands cannot work for massless particles, such as gauge bosons. Assigning a fictitious mass to gauge bosons does not solve the problem either, as such a mass would break gauge invariance. However, this problem occurs in the infrared rather than in the UV. Therefore, we expect that a ``gauge symmetry preserving generalisation" of the CS method does not change our hierarchy discussion. The fact that the 't Hooft method does not seem to have this problem strengthens this intuition. Finally, it seems worth repeating the discussion in the setting of a theory with spontaneous symmetry breaking. It can be shown that there are no fundamental differences in the conclusions \cite{SM2}. That contradicts the common lore \cite{Gildener:1976ai} that radiative corrections set an upper bound on gauge-symmetry hierarchies.

\subsection{New physics}
The discussion of the technical hierarchy problem may sound be quite innocent: after all, different formulations of QFT lead to identical mappings of one set of finite quantities (parameters of the theory) to another set of finite quantities (results of measurements, cross-sections, etc.). However, the conclusions drawn about new physics are very different: ``naturalness''  leads to the conjecture about the existence of new physics right above the Fermi scale\footnote{For example, one of the motivations which were put forward for the construction of the 100 TeV proton-proton collider is the test of the ``naturalness'' of the SM \cite{Strategy:2019vxc}.}, whereas the use of a finite formulation of QFT says that no such a conclusion can be made on physical grounds. At this point it is worth stressing that the aim of our paper was not to formulate a ``postdiction" of the (current) negative results in the search for new TeV physics. We just want to argue that one should not worry about the ``natural stability" of the Fermi scale. 

\subsection{Origin of scales}

Though the problem of the {\em quantum stability} of the Higgs mass and cosmological constant can be resolved by finite QFT, the question about the {\em origin} of widely separated scales in Nature (such as vacuum energy, Fermi, GUT or Planck scale) remains. Refs. \cite{Weinberg:1978ym,Shaposhnikov:2018xkv,Shaposhnikov:2020geh,Shaposhnikov:2009pv,Shaposhnikov:2018jag,Oda:2018zth,Haruna:2019zeu} explore some ideas in this direction, see also below.

\section{Outlook\label{outlook}}

The technical aspect of the hierarchy problem (fine-tuning between the tree values of the bare parameters and radiative corrections in the theories with well-separated mass scales) was the driving force of research in physics beyond the Standard Model for many years. The requirement of the absence of quadratic divergencies was taken as a necessary criterion for avoiding the hierarchy problem (see, e.g. \cite{Martin:1997ns}). The finite approach to QFT, which we developed in our work, tells that the presence of infinities (quartic and quadratic in particular) is formalism dependent, and thus their absence or presence should not be served as an argument for selecting this or that physical theory.  For example, an argument in favour of supersymmetry or composite Higgs models based on the absence of quadratic divergences becomes diluted.

In the models we considered in this work, the essential parameters such as masses of the particles and their couplings cannot be predicted and should be taken from experiments. In the finite formulation of QFT, applied to these theories, no fine-tunings appear at any step of the calculation of different observables. Many popular models of particle physics fall in this class: the Standard Model itself (the analogue of our one-field theory with a single mass scale), non-supersymmetric Grand Unified Theories such as SU(5) or SO(10) (the analogue of our two-field theory with well-separated mass scales), high and low scale see-saw models, Minimal Supersymmetric Standard Model with and without soft SUSY breaking terms,  and SUSY GUTs, to name the few.

The Standard Model so far is consistent with the plentitude of experiments in particle physics, whereas its modest extension by three right-handed neutrinos can solve all its outstanding observational problems - neutrino masses, baryon asymmetry of the Universe, and dark matter (for a review see \cite{Boyarsky:2009ix}). Both theories are self-consistent and are valid as effective field theories till energies much exceeding the Planck scale.  Phenomenologically, three right-handed neutrinos may have masses below the Fermi scale, meaning that the theory consistent with all observations may be characterised just by one essential mass scale, namely that of Fermi. And, therefore, there are no fine-tunings or hierarchy problem in a theory describing all observed phenomena in particle physics.

The only experimental evidence for the presence of another scale in Nature is gravity, meaning that any realistic description of it must have at least two widely separated essential scales, that of Fermi ($M_F \sim 100$ GeV) and Planck ( $M_P \sim 10^{19}$ GeV). Let us discuss first the minimal possibility that there are no other scales. Of course, gravity is not well understood, and the different ideas overviewed below are based on several unproven, though plausible assumptions. 

The Einstein gravity is a non-renormalisable theory, meaning that if perturbation theory is used without any extra input, the SM (or $\nu$MSM) plus gravity is characterised by an infinite number of parameters - couplings of the Standard Model, the Planck mass $M_P$, and coefficients in front of higher order operators, suppressed presumably by $M_P$. Though gravity is non-renormalisable by perturbative methods, it may be UV complete non-perturbatively. 

The first idea -- asymptotic safety --  is due to Weinberg \cite{Weinberg:1980gg}, who argued that the Einstein theory may exhibit a non-trivial ultraviolet fixed point of the functional renormalisation group flow \cite{Wilson:1973jj,Polchinski:1983gv,Wetterich:1992yh}  without any new degrees of freedom. In \cite{Reuter:1996cp} such a fixed point was indeed found in certain approximations. Many works (for a recent review see \cite{Dupuis:2020fhh}) produced further evidence in favour of this conjecture. In \cite{Shaposhnikov:2009pv}  it was proposed to extend the asymptotic safety idea to the combination of the SM and gravity. If the number of parameters characterising the non-trivial UV fixed point is smaller than that of the SM, some of its parameters can be predicted. These considerations led to a successful prediction of the Higgs mass (to be more precise, the low-energy Higgs boson self-coupling, related to the Higgs mass via the Fermi scale),  two years before its discovery at LHC \cite{Shaposhnikov:2009pv}.

The second idea is due to Dvali and Gomez, who argued in \cite{Dvali:2010bf} that Einstein's gravity may be self-complete (meaning that no any extra degrees are needed) due to the existence of black holes. If true, the ultimate theory of Nature may be just the $\nu$MSM plus gravity, with uncomputable parameters including the mass of the Higgs boson, but without any fine-tunings and thus without the technical part of the hierarchy problem. It remains to be seen whether the UV completion of gravity and the Standard Model by black holes may be combined with the asymptotic safety scenario. 

It is an open question whether gravity contributes to the Higgs mass. If $m_H$ is uncomputable, as, in the canonical Standard Model, this question does not make sense, as the mass of the Higgs boson is simply an arbitrary parameter fixed by the experiment. The underlying theory should have fewer parameters than the Standard model to have a framework where the Higgs mass is predictable. 

The idea which attracted a lot of attention is associated with the classical scale invariance and goes back to a seminal paper by Coleman and Weinberg \cite{Coleman:1973jx}.  In these theories, the breaking of the gauge symmetry occurs due to radiative corrections and is associated with scale anomaly. Consequently, the scalar mass can be computed in terms of the vector boson mass. This scenario does not work in the SM with the physical value of the top quark mass (it leads to the Higgs and the top quark masses $m_H \simeq 7$ GeV and  $m_t \lesssim 80$ GeV \cite{Linde:1975sw,Weinberg:1976pe,Linde:1977mm}), but may take place in extended theories (see \cite{Gildener:1976ih}  for initial work).  Clearly,  in addition to the scale anomaly, the theory with gravity has another source of conformal symmetry breaking, associated with the Planck scale. Refs.
\cite{Shaposhnikov:2018jag,Shaposhnikov:2018xkv,Shkerin:2019mmu,Shaposhnikov:2020geh}  explored a certain set of semiclassical configurations which 
can generate in a non-perturbative way the Fermi scale, many orders of magnitudes smaller than the Planck scale, $M_F \propto \exp(-S) M_P$, where $S$ is the classical action of the relevant scalar-gravity configuration.

Let us discuss now the hypothesis that a desert between the Fermi scale and the Planck scale is populated by some other physical scales. The popular candidates are the scale of supersymmetry breaking, the scale of Higgs boson compositeness, and the scale of Grand Unification\footnote{Note that the gauge coupling unification does not necessarily need the existence of particles beyond the Standard Model, see  \cite{Karananas:2017mxm}.}, the Peccei-Quinn scale for axion physics, the scale of inflation, etc. As we have already mentioned, theories of this sort do not have fine-tuning issues if the Higgs mass in them is not computable.

However, one can assume that the field theory in which the Standard Model is embedded is UV complete and contains fewer parameters than the Standard Model. Needless to say that this is a very strong assumption. For example,   gravity enters into the game at energies just a couple of orders of magnitude larger than, say, the scale of Grand Unification, so neglecting it does not sound to be well justified.

To the best of our knowledge, no UV complete field theory that has fewer free parameters than the Standard Model has at least two well-separated mass scales, and agrees with all the data,  has been invented up to now.  Most of the semi-realistic models constructed so far are associated with dynamically broken SUSY with or composite Higgs boson, which use a QCD-like setup. There is no experimental evidence in favour of the generic predictions of these theories.

\section*{Acknowledgements}

We thank  Stan Bentvelsen, Konstantin Chetyrkin, Gia Dvali, Angelo Esposito, Andrei Golutvin, Gerard 't Hooft, Dmitry Kazakov, Marieke Postma, Oleg Ruchayskiy, Bert Schellekens, Elena Shaposhnikova, Jan Smit, Oleksandr Sobol, Inar Timiryasov, the Nikhef theory group, and especially Alexey Boyarsky for useful discussions and comments on the manuscript. In addition to that, we want to thank many listeners for our seminars presenting this work. Their questions, comments and suggestions have been the motivation for the updated version of this paper.

This work was supported by ERC-AdG-2015 grant 694896, by the Swiss National Science
Foundation Excellence grant 200020B\underline{ }182864, and by the Generalitat Valenciana grant PROMETEO/2021/083.  

\bibliographystyle{JHEP2}

\begin{thebibliography}{10}

\bibitem{Gildener:1976ai}
E.~Gildener, \emph{{Gauge Symmetry Hierarchies}},
  \href{https://doi.org/10.1103/PhysRevD.14.1667}{\emph{Phys. Rev. D}
  {\bfseries 14} (1976) 1667}.

\bibitem{Weinberg:1975gm}
S.~Weinberg, \emph{{Implications of Dynamical Symmetry Breaking}},
  \href{https://doi.org/10.1103/PhysRevD.19.1277}{\emph{Phys. Rev. D}
  {\bfseries 13} (1976) 974}.

\bibitem{Buras:1977yy}
A.J.~Buras, J.R.~Ellis, M.K.~Gaillard and D.V.~Nanopoulos, \emph{{Aspects of
  the Grand Unification of Strong, Weak and Electromagnetic Interactions}},
  \href{https://doi.org/10.1016/0550-3213(78)90214-6}{\emph{Nucl. Phys. B}
  {\bfseries 135} (1978) 66}.

\bibitem{Susskind:1978ms}
L.~Susskind, \emph{{Dynamics of Spontaneous Symmetry Breaking in the
  Weinberg-Salam Theory}},
  \href{https://doi.org/10.1103/PhysRevD.20.2619}{\emph{Phys. Rev. D}
  {\bfseries 20} (1979) 2619}.

\bibitem{Susskind:1982mw}
L.~Susskind, \emph{{The Gauge Hierarchy Problem, Technicolor, Supersymmetry,
  and All That}},
  \href{https://doi.org/10.1016/0370-1573(84)90208-4}{\emph{Phys. Rept.}
  {\bfseries 104} (1984) 181}.

\bibitem{Haber:1984rc}
H.E.~Haber and G.L.~Kane, \emph{{The Search for Supersymmetry: Probing Physics
  Beyond the Standard Model}},
  \href{https://doi.org/10.1016/0370-1573(85)90051-1}{\emph{Phys. Rept.}
  {\bfseries 117} (1985) 75}.

\bibitem{Dvali:1995qt}
G.~Dvali, \emph{{Hierarchy problem in SUSY GUTs}},  in \emph{{ICTP Summer
  School in High-energy Physics and Cosmology}}, pp.~605--617, 6, 1995.

\bibitem{Martin:1997ns}
S.P.~Martin, \emph{{A Supersymmetry primer}},
  \href{https://doi.org/10.1142/9789812839657_0001}{\emph{Adv. Ser. Direct.
  High Energy Phys.} {\bfseries 18} (1998) 1}
  [\href{https://arxiv.org/abs/hep-ph/9709356}{{\ttfamily hep-ph/9709356}}].

\bibitem{Chung:2003fi}
D.J.H.~Chung, L.L.~Everett, G.L.~Kane, S.F.~King, J.D.~Lykken and L.-T.~Wang,
  \emph{{The Soft supersymmetry breaking Lagrangian: Theory and applications}},
  \href{https://doi.org/10.1016/j.physrep.2004.08.032}{\emph{Phys. Rept.}
  {\bfseries 407} (2005) 1}
  [\href{https://arxiv.org/abs/hep-ph/0312378}{{\ttfamily hep-ph/0312378}}].

\bibitem{Giudice:2008bi}
G.F.~Giudice, \emph{{Naturally Speaking: The Naturalness Criterion and Physics
  at the LHC}},  \href{https://arxiv.org/abs/0801.2562}{{\ttfamily 0801.2562}}.

\bibitem{Feng:2013pwa}
J.L.~Feng, \emph{{Naturalness and the Status of Supersymmetry}},
  \href{https://doi.org/10.1146/annurev-nucl-102010-130447}{\emph{Ann. Rev.
  Nucl. Part. Sci.} {\bfseries 63} (2013) 351}
  [\href{https://arxiv.org/abs/1302.6587}{{\ttfamily 1302.6587}}].

\bibitem{Dine:2015xga}
M.~Dine, \emph{{Naturalness Under Stress}},
  \href{https://doi.org/10.1146/annurev-nucl-102014-022053}{\emph{Ann. Rev.
  Nucl. Part. Sci.} {\bfseries 65} (2015) 43}
  [\href{https://arxiv.org/abs/1501.01035}{{\ttfamily 1501.01035}}].

\bibitem{Nath:2020xiz}
P.~Nath, \emph{{Supersymmetry unification, naturalness, and discovery prospects
  at HL-LHC and HE-LHC}},
  \href{https://doi.org/10.1140/epjst/e2020-000021-4}{\emph{Eur. Phys. J. ST}
  {\bfseries 229} (2020) 3047}.

\bibitem{Hebecker:2021egx}
A.~Hebecker, \emph{{Naturalness, String Landscape and Multiverse: A Modern
  Introduction with Exercises}}, vol.~979 of \emph{Lecture Notes in Physics}
  (3, 2021),
  \href{https://doi.org/10.1007/978-3-030-65151-0}{10.1007/978-3-030-65151-0}.

\bibitem{Barbieri:1987fn}
R.~Barbieri and G.F.~Giudice, \emph{{Upper Bounds on Supersymmetric Particle
  Masses}}, \href{https://doi.org/10.1016/0550-3213(88)90171-X}{\emph{Nucl.
  Phys. B} {\bfseries 306} (1988) 63}.

\bibitem{Contino:2017moj}
R.~Contino, D.~Greco, R.~Mahbubani, R.~Rattazzi and R.~Torre, \emph{{Precision
  Tests and Fine Tuning in Twin Higgs Models}},
  \href{https://doi.org/10.1103/PhysRevD.96.095036}{\emph{Phys. Rev. D}
  {\bfseries 96} (2017) 095036}
  [\href{https://arxiv.org/abs/1702.00797}{{\ttfamily 1702.00797}}].

\bibitem{deBlas:2019rxi}
J.~de~Blas et~al., \emph{{Higgs Boson Studies at Future Particle Colliders}},
  \href{https://doi.org/10.1007/JHEP01(2020)139}{\emph{JHEP} {\bfseries 01}
  (2020) 139} [\href{https://arxiv.org/abs/1905.03764}{{\ttfamily
  1905.03764}}].

\bibitem{Strategy:2019vxc}
R.K.~Ellis et~al., \emph{{Physics Briefing Book}: {Input for the European
  Strategy for Particle Physics Update 2020}},
  \href{https://arxiv.org/abs/1910.11775}{{\ttfamily 1910.11775}}.

\bibitem{Weinberg:1979sa}
S.~Weinberg, \emph{{Baryon and Lepton Nonconserving Processes}},
  \href{https://doi.org/10.1103/PhysRevLett.43.1566}{\emph{Phys. Rev. Lett.}
  {\bfseries 43} (1979) 1566}.

\bibitem{Wilczek:1979hc}
F.~Wilczek and A.~Zee, \emph{{Operator Analysis of Nucleon Decay}},
  \href{https://doi.org/10.1103/PhysRevLett.43.1571}{\emph{Phys. Rev. Lett.}
  {\bfseries 43} (1979) 1571}.

\bibitem{Asaka:2005pn}
T.~Asaka and M.~Shaposhnikov, \emph{{The $\nu$MSM, dark matter and baryon
  asymmetry of the universe}},
  \href{https://doi.org/10.1016/j.physletb.2005.06.020}{\emph{Phys. Lett. B}
  {\bfseries 620} (2005) 17}
  [\href{https://arxiv.org/abs/hep-ph/0505013}{{\ttfamily hep-ph/0505013}}].

\bibitem{Bezrukov:2007ep}
F.L.~Bezrukov and M.~Shaposhnikov, \emph{{The Standard Model Higgs boson as the
  inflaton}}, \href{https://doi.org/10.1016/j.physletb.2007.11.072}{\emph{Phys.
  Lett. B} {\bfseries 659} (2008) 703}
  [\href{https://arxiv.org/abs/0710.3755}{{\ttfamily 0710.3755}}].

\bibitem{dva3}
G.~Dvali, \emph{{Strong Coupling and Classicalization}},
  \href{https://doi.org/10.1142/9789813208292_0005}{\emph{Subnucl. Ser.}
  {\bfseries 53} (2017) 189}
  [\href{https://arxiv.org/abs/1607.07422}{{\ttfamily 1607.07422}}].

\bibitem{Georgi:1974yf}
H.~Georgi, H.R.~Quinn and S.~Weinberg, \emph{{Hierarchy of Interactions in
  Unified Gauge Theories}},
  \href{https://doi.org/10.1103/PhysRevLett.33.451}{\emph{Phys. Rev. Lett.}
  {\bfseries 33} (1974) 451}.

\bibitem{hossi}
S.~Hossenfelder, \emph{{Screams for explanation: finetuning and naturalness in
  the foundations of physics}},
  \href{https://doi.org/10.1007/s11229-019-02377-5}{\emph{Synthese} {\bfseries
  198} (2021) 3727} [\href{https://arxiv.org/abs/1801.02176}{{\ttfamily
  1801.02176}}].

\bibitem{wells1}
J.D.~Wells, \emph{{Naturalness, Extra-Empirical Theory Assessments, and the
  Implications of Skepticism}},
  \href{https://doi.org/10.1007/s10701-018-0220-x}{\emph{Found. Phys.}
  {\bfseries 49} (2019) 991}
  [\href{https://arxiv.org/abs/1806.07289}{{\ttfamily 1806.07289}}].

\bibitem{wells2}
J.D.~Wells, \emph{{Finetuned Cancellations and Improbable Theories}},
  \href{https://doi.org/10.1007/s10701-019-00254-2}{\emph{Found. Phys.}
  {\bfseries 49} (2019) 428}
  [\href{https://arxiv.org/abs/1809.03374}{{\ttfamily 1809.03374}}].

\bibitem{Ellis:2021kzk}
J.~Ellis, \emph{{SMEFT Constraints on New Physics Beyond the Standard Model}},
  in \emph{{Beyond Standard Model: From Theory to Experiment}}, 5, 2021
  [\href{https://arxiv.org/abs/2105.14942}{{\ttfamily 2105.14942}}].

\bibitem{Brivio:2017vri}
I.~Brivio and M.~Trott, \emph{{The Standard Model as an Effective Field
  Theory}}, \href{https://doi.org/10.1016/j.physrep.2018.11.002}{\emph{Phys.
  Rept.} {\bfseries 793} (2019) 1}
  [\href{https://arxiv.org/abs/1706.08945}{{\ttfamily 1706.08945}}].

\bibitem{LHCb:2021trn}
{\scshape LHCb} collaboration, \emph{{Test of lepton universality in
  beauty-quark decays}},  \href{https://arxiv.org/abs/2103.11769}{{\ttfamily
  2103.11769}}.

\bibitem{Muong-2:2021ojo}
{\scshape Muon g-2} collaboration, \emph{{Measurement of the Positive Muon
  Anomalous Magnetic Moment to 0.46 ppm}},
  \href{https://doi.org/10.1103/PhysRevLett.126.141801}{\emph{Phys. Rev. Lett.}
  {\bfseries 126} (2021) 141801}
  [\href{https://arxiv.org/abs/2104.03281}{{\ttfamily 2104.03281}}].

\bibitem{Giudice:2013yca}
G.F.~Giudice, \emph{{Naturalness after LHC8}},
  \href{https://doi.org/10.22323/1.180.0163}{\emph{PoS} {\bfseries EPS-HEP2013}
  (2013) 163} [\href{https://arxiv.org/abs/1307.7879}{{\ttfamily 1307.7879}}].

\bibitem{ParticleDataGroup:2020ssz}
{\scshape Particle Data Group} collaboration, \emph{{Review of Particle
  Physics}}, \href{https://doi.org/10.1093/ptep/ptaa104}{\emph{PTEP} {\bfseries
  2020} (2020) 083C01}.

\bibitem{dva1}
G.~Dvali and A.~Vilenkin, \emph{{Cosmic attractors and gauge hierarchy}},
  \href{https://doi.org/10.1103/PhysRevD.70.063501}{\emph{Phys. Rev. D}
  {\bfseries 70} (2004) 063501}
  [\href{https://arxiv.org/abs/hep-th/0304043}{{\ttfamily hep-th/0304043}}].

\bibitem{dva2}
G.~Dvali, \emph{{Large hierarchies from attractor vacua}},
  \href{https://doi.org/10.1103/PhysRevD.74.025018}{\emph{Phys. Rev. D}
  {\bfseries 74} (2006) 025018}
  [\href{https://arxiv.org/abs/hep-th/0410286}{{\ttfamily hep-th/0410286}}].

\bibitem{Graham:2015cka}
P.W.~Graham, D.E.~Kaplan and S.~Rajendran, \emph{{Cosmological Relaxation of
  the Electroweak Scale}},
  \href{https://doi.org/10.1103/PhysRevLett.115.221801}{\emph{Phys. Rev. Lett.}
  {\bfseries 115} (2015) 221801}
  [\href{https://arxiv.org/abs/1504.07551}{{\ttfamily 1504.07551}}].

\bibitem{Giudice:2021viw}
G.F.~Giudice, M.~McCullough and T.~You, \emph{{Self-Organised Localisation}},
  \href{https://arxiv.org/abs/2105.08617}{{\ttfamily 2105.08617}}.

\bibitem{Banerjee:2020kww}
A.~Banerjee, H.~Kim, O.~Matsedonskyi, G.~Perez and M.S.~Safronova,
  \emph{{Probing the Relaxed Relaxion at the Luminosity and Precision
  Frontiers}}, \href{https://doi.org/10.1007/JHEP07(2020)153}{\emph{JHEP}
  {\bfseries 07} (2020) 153}
  [\href{https://arxiv.org/abs/2004.02899}{{\ttfamily 2004.02899}}].

\bibitem{Cheung:2017pzi}
C.~Cheung, \emph{{TASI Lectures on Scattering Amplitudes}},  in
  \emph{{Proceedings, Theoretical Advanced Study Institute in Elementary
  Particle Physics : Anticipating the Next Discoveries in Particle Physics
  (TASI 2016)}: {Boulder, CO, USA, June 6-July 1, 2016}}, R.~Essig and I.~Low,
  eds. (2018), \href{https://doi.org/10.1142/9789813233348\_0008}{DOI}
  [\href{https://arxiv.org/abs/1708.03872}{{\ttfamily 1708.03872}}].

\bibitem{Rattazzi:2008pe}
R.~Rattazzi, V.S.~Rychkov, E.~Tonni and A.~Vichi, \emph{{Bounding scalar
  operator dimensions in 4D CFT}},
  \href{https://doi.org/10.1088/1126-6708/2008/12/031}{\emph{JHEP} {\bfseries
  12} (2008) 031} [\href{https://arxiv.org/abs/0807.0004}{{\ttfamily
  0807.0004}}].

\bibitem{El-Showk:2012cjh}
S.~El-Showk, M.F.~Paulos, D.~Poland, S.~Rychkov, D.~Simmons-Duffin and
  A.~Vichi, \emph{{Solving the 3D Ising Model with the Conformal Bootstrap}},
  \href{https://doi.org/10.1103/PhysRevD.86.025022}{\emph{Phys. Rev. D}
  {\bfseries 86} (2012) 025022}
  [\href{https://arxiv.org/abs/1203.6064}{{\ttfamily 1203.6064}}].

\bibitem{Paulos:2016fap}
M.F.~Paulos, J.~Penedones, J.~Toledo, B.C.~van Rees and P.~Vieira, \emph{{The
  S-matrix bootstrap. Part I: QFT in AdS}},
  \href{https://doi.org/10.1007/JHEP11(2017)133}{\emph{JHEP} {\bfseries 11}
  (2017) 133} [\href{https://arxiv.org/abs/1607.06109}{{\ttfamily
  1607.06109}}].

\bibitem{Collins:1984xc}
J.C.~Collins, \emph{{Renormalization}: {An Introduction to Renormalization, The
  Renormalization Group, and the Operator Product Expansion}}, vol.~26 of
  \emph{Cambridge Monographs on Mathematical Physics}, Cambridge University
  Press, Cambridge (1986),
  \href{https://doi.org/10.1017/CBO9780511622656}{10.1017/CBO9780511622656}.

\bibitem{Bogoliubov}
N.~Bogoliubov and D.~Shirkov, \emph{{Introduction to Theory of Quantized
  Fields}}, John Wiley \& Sons Inc (1959).

\bibitem{Kazakov:2020mfp}
D.I.~Kazakov, \emph{{The Bogolyubov $\mathcal{R}$-Operation in
  Nonrenormalizable Theories}},
  \href{https://doi.org/10.1134/S1063779620040383}{\emph{Phys. Part. Nucl.}
  {\bfseries 51} (2020) 503}.

\bibitem{Lenshina:2020edt}
N.D.~Lenshina, A.A.~Radionov and F.V.~Tkachov, \emph{{MS$^{4}$: An Alternative
  to the
  Bogolyubov\textendash{}Parasiuk\textendash{}Hepp\textendash{}Zimmermann
  (BPHZ) Theory}}, \href{https://doi.org/10.1134/S1063779620040462}{\emph{Phys.
  Part. Nucl.} {\bfseries 51} (2020) 567}.

\bibitem{Callan:1970yg}
C.G.~Callan, Jr., \emph{{Broken scale invariance in scalar field theory}},
  \href{https://doi.org/10.1103/PhysRevD.2.1541}{\emph{Phys. Rev. D} {\bfseries
  2} (1970) 1541}.

\bibitem{Symanzik:1970rt}
K.~Symanzik, \emph{{Small distance behavior in field theory and power
  counting}}, \href{https://doi.org/10.1007/BF01649434}{\emph{Commun. Math.
  Phys.} {\bfseries 18} (1970) 227}.

\bibitem{Blaer:1974foy}
A.S.~Blaer and K.~Young, \emph{{Field theory renormalization using the
  Callan-Symanzik equation}},
  \href{https://doi.org/10.1016/0550-3213(74)90271-5}{\emph{Nucl. Phys. B}
  {\bfseries 83} (1974) 493}.

\bibitem{Callan:1975vs}
C.G.~Callan, Jr., \emph{{Introduction to Renormalization Theory}}, {\emph{Conf.
  Proc. C} {\bfseries 7507281} (1975) 41}.

\bibitem{tHooft:2004bkn}
G.~'t~Hooft, \emph{{Renormalization without infinities}},
  \href{https://doi.org/10.1142/S0217751X05024249}{\emph{Int. J. Mod. Phys. A}
  {\bfseries 20} (2005) 1336}
  [\href{https://arxiv.org/abs/hep-th/0405032}{{\ttfamily hep-th/0405032}}].

\bibitem{Lehmann:1954rq}
H.~Lehmann, K.~Symanzik and W.~Zimmermann, \emph{{On the formulation of
  quantized field theories}},
  \href{https://doi.org/10.1007/BF02731765}{\emph{Nuovo Cim.} {\bfseries 1}
  (1955) 205}.

\bibitem{Nishijima:1960zz}
K.~Nishijima, \emph{{Asymptotic Conditions and Perturbation Theory}},
  \href{https://doi.org/10.1103/PhysRev.119.485}{\emph{Phys. Rev.} {\bfseries
  119} (1960) 485}.

\bibitem{Kasia}
K.~Rejzner, \emph{{Perturbative Algebraic Quantum Field Theory}: {An
  Introduction for Mathematicians}}, Mathematical Physics Studies, Springer,
  New York (2016),
  \href{https://doi.org/10.1007/978-3-319-25901-7}{10.1007/978-3-319-25901-7}.

\bibitem{Morgan}
P.~Morgan, \emph{{A source fragmentation approach to interacting quantum field
  theory}},  \href{https://arxiv.org/abs/2109.04412}{{\ttfamily 2109.04412}}.

\bibitem{moff1}
J.W.~Moffat, \emph{{Ultraviolet Complete Quantum Gravity}},
  \href{https://doi.org/10.1140/epjp/i2011-11043-7}{\emph{Eur. Phys. J. Plus}
  {\bfseries 126} (2011) 43} [\href{https://arxiv.org/abs/1008.2482}{{\ttfamily
  1008.2482}}].

\bibitem{moff2}
J.W.~Moffat, \emph{{Quantum Gravity and the Cosmological Constant Problem}},
  \href{https://doi.org/10.1007/978-3-319-20046-0_36}{\emph{Springer Proc.
  Phys.} {\bfseries 170} (2016) 299}
  [\href{https://arxiv.org/abs/1407.2086}{{\ttfamily 1407.2086}}].

\bibitem{moff3}
J.W.~Moffat, \emph{{Ultraviolet Complete Quantum Field Theory and Particle
  Model}}, \href{https://doi.org/10.1140/epjp/i2019-12973-6}{\emph{Eur. Phys.
  J. Plus} {\bfseries 134} (2019) 443}
  [\href{https://arxiv.org/abs/1812.01986}{{\ttfamily 1812.01986}}].

\bibitem{moff4}
J.W.~Moffat, \emph{{Model of Boson and Fermion Particle Masses}},
  \href{https://doi.org/10.1140/epjp/s13360-021-01608-4}{\emph{Eur. Phys. J.
  Plus} {\bfseries 136} (2021) 601}
  [\href{https://arxiv.org/abs/2009.10145}{{\ttfamily 2009.10145}}].

\bibitem{moff5}
M.A.~Green and J.W.~Moffat, \emph{{Finite Quantum Field Theory and
  Renormalization Group}}, {\emph{Eur. Phys. J. Plus} {\bfseries 131} (2021)
  919} [\href{https://arxiv.org/abs/2012.04487}{{\ttfamily 2012.04487}}].

\bibitem{SanderMisha1}
S.~Mooij and M.~Shaposhnikov, \emph{{Finite Callan-Symanzik renormalisation for
  multiple scalar fields}},  \href{https://doi.org/10.1016/j.nuclphysb.2023.116176}{\emph{Nucl. Phys. B} {\bfseries 990}
  (2023) 116176}  [\href{https://arxiv.org/abs/2110.15925}{{\ttfamily
  2110.15925}}].


\bibitem{Peskin:1995ev}
M.E.~Peskin and D.V.~Schroeder, \emph{{An Introduction to quantum field
  theory}}, Addison-Wesley, Reading, USA (1995).

\bibitem{Naud:1998yg}
J.~Naud, I.~Nemenman, M.~Van~Raamsdonk and V.~Periwal, \emph{{Minimal
  subtraction and the Callan-Symanzik equation}},
  \href{https://doi.org/10.1016/S0550-3213(98)00665-8}{\emph{Nucl. Phys. B}
  {\bfseries 540} (1999) 533}
  [\href{https://arxiv.org/abs/hep-th/9802181}{{\ttfamily hep-th/9802181}}].

\bibitem{SM2}
S.~Mooij and M.~Shaposhnikov, \emph{{in preparation}}, .

\bibitem{Wess:1992cp}
J.~Wess and J.~Bagger, \emph{{Supersymmetry and supergravity}}, Princeton
  University Press, Princeton, NJ, USA (1992).

\bibitem{Manohar:2018aog}
A.V.~Manohar, \emph{{Introduction to Effective Field Theories}},
  \href{https://arxiv.org/abs/1804.05863}{{\ttfamily 1804.05863}}.

\bibitem{Shaposhnikov:2007nj}
M.~Shaposhnikov, \emph{{Is there a new physics between electroweak and Planck
  scales?}},  in \emph{{Astroparticle Physics: Current Issues, 2007 (APCI07)}},
  8, 2007 [\href{https://arxiv.org/abs/0708.3550}{{\ttfamily 0708.3550}}].

\bibitem{Vissani:1997ys}
F.~Vissani, \emph{{Do experiments suggest a hierarchy problem?}},
  \href{https://doi.org/10.1103/PhysRevD.57.7027}{\emph{Phys. Rev. D}
  {\bfseries 57} (1998) 7027}
  [\href{https://arxiv.org/abs/hep-ph/9709409}{{\ttfamily hep-ph/9709409}}].

\bibitem{Farina:2013mla}
M.~Farina, D.~Pappadopulo and A.~Strumia, \emph{{A modified naturalness
  principle and its experimental tests}},
  \href{https://doi.org/10.1007/JHEP08(2013)022}{\emph{JHEP} {\bfseries 08}
  (2013) 022} [\href{https://arxiv.org/abs/1303.7244}{{\ttfamily 1303.7244}}].

\bibitem{Cohen:2019wxr}
T.~Cohen, \emph{{As Scales Become Separated: Lectures on Effective Field
  Theory}}, {\emph{PoS} {\bfseries TASI2018} (2019) 011}
  [\href{https://arxiv.org/abs/1903.03622}{{\ttfamily 1903.03622}}].

\bibitem{Gaillard}
M.K.~Gaillard and B.W.~Lee, \emph{{Rare Decay Modes of the K-Mesons in Gauge
  Theories}}, \href{https://doi.org/10.1103/PhysRevD.10.897}{\emph{Phys. Rev.
  D} {\bfseries 10} (1974) 897}.

\bibitem{Das}
T.~Das, G.S.~Guralnik, V.S.~Mathur, F.E.~Low and J.E.~Young,
  \emph{{Electromagnetic mass difference of pions}},
  \href{https://doi.org/10.1103/PhysRevLett.18.759}{\emph{Phys. Rev. Lett.}
  {\bfseries 18} (1967) 759}.

\bibitem{Coleman:1973jx}
S.R.~Coleman and E.J.~Weinberg, \emph{{Radiative Corrections as the Origin of
  Spontaneous Symmetry Breaking}},
  \href{https://doi.org/10.1103/PhysRevD.7.1888}{\emph{Phys. Rev. D} {\bfseries
  7} (1973) 1888}.

\bibitem{Weinberg:1978ym}
S.~Weinberg, \emph{{Gauge Hierarchies}},
  \href{https://doi.org/10.1016/0370-2693(79)90248-X}{\emph{Phys. Lett. B}
  {\bfseries 82} (1979) 387}.

\bibitem{Shaposhnikov:2018xkv}
M.~Shaposhnikov and A.~Shkerin, \emph{{Conformal symmetry: towards the link
  between the Fermi and the Planck scales}},
  \href{https://doi.org/10.1016/j.physletb.2018.06.068}{\emph{Phys. Lett. B}
  {\bfseries 783} (2018) 253}
  [\href{https://arxiv.org/abs/1803.08907}{{\ttfamily 1803.08907}}].

\bibitem{Shaposhnikov:2020geh}
M.~Shaposhnikov, A.~Shkerin and S.~Zell, \emph{{Standard Model Meets Gravity:
  Electroweak Symmetry Breaking and Inflation}},
  \href{https://doi.org/10.1103/PhysRevD.103.033006}{\emph{Phys. Rev. D}
  {\bfseries 103} (2021) 033006}
  [\href{https://arxiv.org/abs/2001.09088}{{\ttfamily 2001.09088}}].

\bibitem{Appelquist:1974tg}
T.~Appelquist and J.~Carazzone, \emph{{Infrared Singularities and Massive
  Fields}}, \href{https://doi.org/10.1103/PhysRevD.11.2856}{\emph{Phys. Rev. D}
  {\bfseries 11} (1975) 2856}.

\bibitem{Neumaier}
A.~Neumaier, \emph{{Renormalization without infinities - an elementary
  tutorial}}, {\emph{\url{https://www.mat.univie.ac.at/~neum/ms/ren.pdf}}
  (2015) }.

\bibitem{Shaposhnikov:2009pv}
M.~Shaposhnikov and C.~Wetterich, \emph{{Asymptotic safety of gravity and the
  Higgs boson mass}},
  \href{https://doi.org/10.1016/j.physletb.2009.12.022}{\emph{Phys. Lett. B}
  {\bfseries 683} (2010) 196}
  [\href{https://arxiv.org/abs/0912.0208}{{\ttfamily 0912.0208}}].

\bibitem{Shaposhnikov:2018jag}
M.~Shaposhnikov and A.~Shkerin, \emph{{Gravity, Scale Invariance and the
  Hierarchy Problem}},
  \href{https://doi.org/10.1007/JHEP10(2018)024}{\emph{JHEP} {\bfseries 10}
  (2018) 024} [\href{https://arxiv.org/abs/1804.06376}{{\ttfamily
  1804.06376}}].

\bibitem{Oda:2018zth}
I.~Oda, \emph{{Planck and Electroweak Scales Emerging from Conformal Gravity}},
  \href{https://doi.org/10.1140/epjc/s10052-018-6289-8}{\emph{Eur. Phys. J. C}
  {\bfseries 78} (2018) 798}
  [\href{https://arxiv.org/abs/1806.03420}{{\ttfamily 1806.03420}}].

\bibitem{Haruna:2019zeu}
J.~Haruna and H.~Kawai, \emph{{Weak scale from Planck scale: Mass scale
  generation in a classically conformal two-scalar system}},
  \href{https://doi.org/10.1093/ptep/ptz165}{\emph{PTEP} {\bfseries 2020}
  (2020) 033B01} [\href{https://arxiv.org/abs/1905.05656}{{\ttfamily
  1905.05656}}].

\bibitem{Boyarsky:2009ix}
A.~Boyarsky, O.~Ruchayskiy and M.~Shaposhnikov, \emph{{The Role of sterile
  neutrinos in cosmology and astrophysics}},
  \href{https://doi.org/10.1146/annurev.nucl.010909.083654}{\emph{Ann. Rev.
  Nucl. Part. Sci.} {\bfseries 59} (2009) 191}
  [\href{https://arxiv.org/abs/0901.0011}{{\ttfamily 0901.0011}}].

\bibitem{Weinberg:1980gg}
S.~Weinberg, \emph{{ULTRAVIOLET DIVERGENCES IN QUANTUM THEORIES OF
  GRAVITATION}},  in \emph{{General Relativity}: {An Einstein Centenary
  Survey}}, pp.~790--831 (1980).

\bibitem{Wilson:1973jj}
K.G.~Wilson and J.B.~Kogut, \emph{{The Renormalization group and the epsilon
  expansion}}, \href{https://doi.org/10.1016/0370-1573(74)90023-4}{\emph{Phys.
  Rept.} {\bfseries 12} (1974) 75}.

\bibitem{Polchinski:1983gv}
J.~Polchinski, \emph{{Renormalization and Effective Lagrangians}},
  \href{https://doi.org/10.1016/0550-3213(84)90287-6}{\emph{Nucl. Phys. B}
  {\bfseries 231} (1984) 269}.

\bibitem{Wetterich:1992yh}
C.~Wetterich, \emph{{Exact evolution equation for the effective potential}},
  \href{https://doi.org/10.1016/0370-2693(93)90726-X}{\emph{Phys. Lett. B}
  {\bfseries 301} (1993) 90}
  [\href{https://arxiv.org/abs/1710.05815}{{\ttfamily 1710.05815}}].

\bibitem{Reuter:1996cp}
M.~Reuter, \emph{{Nonperturbative evolution equation for quantum gravity}},
  \href{https://doi.org/10.1103/PhysRevD.57.971}{\emph{Phys. Rev. D} {\bfseries
  57} (1998) 971} [\href{https://arxiv.org/abs/hep-th/9605030}{{\ttfamily
  hep-th/9605030}}].

\bibitem{Dupuis:2020fhh}
N.~Dupuis, L.~Canet, A.~Eichhorn, W.~Metzner, J.M.~Pawlowski, M.~Tissier
  et~al., \emph{{The nonperturbative functional renormalization group and its
  applications}},
  \href{https://doi.org/10.1016/j.physrep.2021.01.001}{\emph{Phys. Rept.}
  {\bfseries 910} (2021) 1} [\href{https://arxiv.org/abs/2006.04853}{{\ttfamily
  2006.04853}}].

\bibitem{Dvali:2010bf}
G.~Dvali and C.~Gomez, \emph{{Self-Completeness of Einstein Gravity}},
  \href{https://arxiv.org/abs/1005.3497}{{\ttfamily 1005.3497}}.

\bibitem{Linde:1975sw}
A.D.~Linde, \emph{{Dynamical Symmetry Restoration and Constraints on Masses and
  Coupling Constants in Gauge Theories}}, {\emph{JETP Lett.} {\bfseries 23}
  (1976) 64}.

\bibitem{Weinberg:1976pe}
S.~Weinberg, \emph{{Mass of the Higgs Boson}},
  \href{https://doi.org/10.1103/PhysRevLett.36.294}{\emph{Phys. Rev. Lett.}
  {\bfseries 36} (1976) 294}.

\bibitem{Linde:1977mm}
A.D.~Linde, \emph{{On the Vacuum Instability and the Higgs Meson Mass}},
  \href{https://doi.org/10.1016/0370-2693(77)90664-5}{\emph{Phys. Lett. B}
  {\bfseries 70} (1977) 306}.

\bibitem{Gildener:1976ih}
E.~Gildener and S.~Weinberg, \emph{{Symmetry Breaking and Scalar Bosons}},
  \href{https://doi.org/10.1103/PhysRevD.13.3333}{\emph{Phys. Rev. D}
  {\bfseries 13} (1976) 3333}.

\bibitem{Shkerin:2019mmu}
A.~Shkerin, \emph{{Dilaton-assisted generation of the Fermi scale from the
  Planck scale}}, \href{https://doi.org/10.1103/PhysRevD.99.115018}{\emph{Phys.
  Rev. D} {\bfseries 99} (2019) 115018}
  [\href{https://arxiv.org/abs/1903.11317}{{\ttfamily 1903.11317}}].

\bibitem{Karananas:2017mxm}
G.K.~Karananas and M.~Shaposhnikov, \emph{{Gauge coupling unification without
  leptoquarks}},
  \href{https://doi.org/10.1016/j.physletb.2017.05.065}{\emph{Phys. Lett. B}
  {\bfseries 771} (2017) 332}
  [\href{https://arxiv.org/abs/1703.02964}{{\ttfamily 1703.02964}}].

\end{thebibliography}

\providecommand{\href}[2]{#2}\begingroup\raggedright\endgroup

\end{document}